\documentclass[useAMS, usenatbib]{mn2e}

\pdfoutput=1
\usepackage{gensymb,upgreek}
\usepackage{amssymb}
\usepackage{graphicx}
\usepackage{epstopdf}
\usepackage{tablefootnote}

\newcommand{\lsun}{\mbox{L$_{\odot}$}}
\newcommand{\msun}{\mbox{M$_{\odot}$}}
\newcommand{\kms}{\mbox{km s$^{-1}$}}

\begin{document}

\title[Gone without a bang]{Gone without a bang: An archival HST survey for disappearing massive stars}

\author[Reynolds, Fraser \& Gilmore]{Thomas M. Reynolds$^{1,2}$\thanks{thomas.reynolds@cantab.net},
    Morgan Fraser$^1$ \&
    Gerard Gilmore$^1$. \\ 
$^1$Institute of Astronomy, University of Cambridge, Madingley Road, Cambridge, CB3 0HA, UK\\
$^2$Tuorla Observatory, Department of Physics and Astronomy, University of Turku, V\"ais\"al\"antie 20, FI-21500 Piikki\"o, Finland\\}

\maketitle
\begin{abstract}
It has been argued that a substantial fraction of massive stars may end their lives without an optically bright supernova (SN), but rather collapse to form a black hole. Such an event would not be detected by current SN surveys, which are focused on finding bright transients. \cite{Koc08} proposed a novel survey for such events, using repeated observations of nearby galaxies to search for the disappearance of a massive star. We present such a survey, using the first systematic analysis of archival {\it Hubble Space Telescope} images of nearby galaxies with the aim of identifying evolved massive stars which have disappeared, {\it without} an accompanying optically bright supernova. We consider a sample of 15 galaxies, with at least three epochs of {\it Hubble Space Telescope} imaging taken between 1994 and 2013. Within this data, we find one candidate which is consistent with a 25--30~\msun\ yellow supergiant which has undergone an optically dark core-collapse. 
\end{abstract}


\section{Introduction}

Massive ($>$8~\msun) stars end their lives as core-collapse supernovae (CCSNe). Once they have evolved off the main sequence and developed a Chandrasekhar-mass Fe core, the pressure in their core is no longer sufficient to support the star against its own gravity. The inner core begins to infall, a process which is halted once it reaches nuclear densities and forms a proto-neutron star (PNS). At this point the core rebounds, and drives a shock outwards through the star. According to current simulations, this shock lacks the energy to halt the stars collapse and explode the outer layers of the star. It is believed that the deposition of additional energy behind the shock (perhaps in the form of neutrinos) can revive it and cause the star to explode as a CCSNe \cite[e.g.][]{Bur13,Jan07}. It is possible, however, for a supernova (SN) explosion to fail. If the energy deposition is insufficient to power the shock to overcome the infalling material, then the shock is not revived \cite[e.g.][]{Woo93,Fry99,Oco11}. This is a more likely outcome for more massive stars, which have bigger Fe cores and are consequently harder to explode. 


SNe are distinguished primarily by the elements seen in their spectra - Type I SNe are H poor, while Type II SNe are H rich \citep{Fil97,Sma09b}. The Type I SNe are further subdivided into Type Ib and Type Ic SNe which show the presence or absence of He respectively. H rich Type II SNe are separated according to their lightcurves into Type IIP (Plateau) SNe, which show an extended period of roughly constant luminosity after explosion, and Type IIL (Linear) SNe which have a steady decline in luminosity from peak. These differing SN types have long been believed to result from the extent of the H and/or He envelope of the progenitor star at the moment of collapse \citep{Fal77}, although recently there has been some debate in the literature as to whether Type IIP and IIL SNe are distinct classes \citep[e.g.][and references therein]{And14,Far14}.

Nearby CCSNe present a unique opportunity to test this hypothesis by directly identifying their massive stellar progenitors in archival images. This was first accomplished for SN 1987A in the Large Magellanic Cloud \citep{Wes87}, and for SN 1993J in M51 \citep{Ald94,Mau09}. Since then, there has been significant success in detecting the H rich progenitors of Type II SNe \cite[e.g.][]{Li07,Mat08,Mau11,Van12,Fra14}. In particular, Type IIP SNe have been shown to come from red supergiants (RSGs) with extended atmospheres \citep{Sma09a}. The H-poor progenitors of Type Ibc SNe, however, have thus far eluded confirmed detections \citep{Eld13}, with only iPTF13bvn \citep{Cao13,Ber14,Eld15} as a viable, albeit unconfirmed, candidate.

While statistical studies of ensembles of SN progenitors \citep{Van03,Sma09a} have revealed that Type IIP SNe do indeed come from RSGs, there is an apparent lack of progenitors with a luminosity comparable to that of the brightest known RSGs. \citeauthor{Sma09a} found that there was a statistically significant absence of higher mass ($\gtrsim$16 \msun) red supergiants exploding as SNe, terming this the ``Red Supergiant Problem''. \citeauthor{Sma09a} went on to suggest that the red supergiant problem might be explained by failed SNe, with stars in the mass range from 16 \msun\ to 30 \msun\ collapsing to form black holes with very weak (or perhaps non-existent) explosions. 

Furthermore, attempts have been made to constrain SN progenitor masses without direct progenitor observations.  By analysing the resolved stellar population in the vicinity of nearby SNe, the local stellar age can be determined, and used to set an upper mass limit on the SN progenitor \citep{Wil14}. \citeauthor{Wil14} set 11 additional (to those of \citep{Sma09a}) progenitor constraints with this method with results consistent with there being no progenitors with $\gtrsim$20 \msun.

There is some supporting evidence for the suggestion that $\gtrsim$16 \msun\ stars do not explode as optically bright SNe. Spectra of SNe taken $\sim$1 yr after explosion can provide a diagnostic of the core mass and composition of the star that exploded \citep{Maz10,Jer12}. In particular, the ejected mass of O, and the strength of the [O~{\sc i}] emission lines increase rapidly for a progenitor with a mass of more than $\sim$20 \msun\ \citep{Jer14}. To date, however, there has been no observed Type IIP SNe with strong [O~{\sc i}] emission consistent with a progenitor with a zero-age main sequence (ZAMS) mass $\gtrsim$16 \msun\ \citep{Jer14}. We also note that \cite{Bro13} found that Galactic abundances can be reproduced even if no stars above 25 \msun\ contribute metals via SN explosions, and that even an even lower threshold of 18 \msun\ could be accommodated with some modification of uncertain mass loss parameters and reaction rates. 

Attempts have also been made to compare the observed SN rate with the star formation rate (SFR) \citep{Hor11,Bot12}. \citeauthor{Hor11} claimed that SN rates were lower than expected given the observed star formation rates in galaxies locally and at intermediate redshifts, and suggested that this may be explained by a significant number of optically dark (either intrinsically or due to dust obscuration) SNe. However, measurements of both the SN rate and SFR suffer from signicant systematic and random uncertainties \citep{Mat14}, and \cite{Bot12} find that there is no disagreement between the SN rate and SFR within 11 Mpc. \cite{Mat12} also find the SN rate and SFR to be in agreement in the very local ($<$15 Mpc) volume but further argue that a significant number of SNe are missed in optical searches beyond the local universe due to obscuration by large amounts of dust in their host galaxies, especially in luminous and ultraluminous infrared galaxies. Studies including corrections for the SNe lost in such highly dust-enshrouded environments have found their rates to be well consistent with the SFR up to z$\sim$1.3 as expected \citep{Dah12,Mel12}.

The relative rates of different CCSN types can also be used to probe their progenitor channel. \cite{Smi11} found that 48 per cent of observed CCSNe were of Type IIP, and that if these came from the lower extremum of the core-collapse mass range, that RSGs between 8.5 and 13.7 \msun\ could account for these. However, \citeauthor{Smi11} went on to argue that the ``missing'' higher mass RSGs may simply continue to evolve into a post-RSG phase, before exploding as a different type of progenitor. A final piece of evidence for the existence of missing SNe lies in the observed distribution of compact remnant masses, where \cite{Koc14} argued that the discontinuity between the masses of neutron stars and black holes could be explained by failed SNe.

Archival searches for the progenitors of CCSNe such as that discussed by \cite{Sma09b}, rely on surveys which are searching for bright optical transients. While \cite{Lov13} and \cite{Pir13} have suggested that a failed SN may be accompanied by a faint transient, to date no convincing candidate has been found. If a core-collapse is optically dark (or very faint), then it will not be detected by these surveys, and hence its progenitor will not be identified (although we note that \cite{Yan11} have raised the possibility of detecting failed SNe through their neutrino emission). As an alternative approach, \cite{Koc08} proposed a ``survey about nothing'' to monitor a sample of nearby galaxies with sufficient depth and resolution to detect individual massive stars. By searching for a massive star which had disappeared, it would be possible to identify faint or dark failed supernovae, which show no bright optical display. \cite{Koc08} suggested either using the {\it Hubble Space Telescope} ({\it HST}) or ground-based 8-m class telescopes for such a survey. \cite{Ger14} presented the first data from such a survey, using a sample of 27 galaxies, observed over 4 years with the Large Binocular Telescope. \citeauthor{Ger14} found a single candidate in their data, which they suggested was consistent with a failed SN fraction of $\sim$0.3 of all core-collapses.

In this paper, we present the first attempt to conduct a systematic search for failed supernovae using extant archival HST data. In Sect. \ref{sect:methods} we discuss the selection criteria for our sample of galaxies, and the technique used to search for disappearing stars; in Sect. \ref{sect:candidates} we present the candidate failed SNe found, while in Sections \ref{sect:discussion} and \ref{sect:conclusions} we provide a discussion of the results and conclusions respectively.

\section{Sample and methodology}
\label{sect:methods}

The observational signature of an optically dark core-collapse SN is a massive star which is seen to disappear without an accompanying bright SN. Our basic approach is to take galaxies which have been observed with the {\it Hubble Space Telescope} ({\it HST}) in the same filter, on at least three separate occasions, and search for any luminous point source which is present in the first two epochs, but is no longer present in the third epoch. In the following section we elaborate on our methodology, and describe in detail the construction of our sample of galaxies and the selection of data used.

\subsection{Target and data selection}

The NASA Extragalactic Database (NED)\footnote{http://ned.ipac.caltech.edu/} was queried for all objects classed as a galaxy; which were listed in either the Messier catalog, the New General Catalog, the Index Catalog, or the Upsala General Catalog of Galaxies; and which had a recessional velocity of less than 2000 \kms. From this, we obtained a catalog of 2665 galaxies and their coordinates, which served as our input sample.

We adopt a maximum distance limit of 28 Mpc (corresponding to a recessional velocity of $<$2000 \kms\ for H$_0$~=~72 km~s$^{-1}$~Mpc$^{-1}$, after correcting for infall on the Virgo cluster) for our sample. This is the same limit adopted by \cite{Sma09a} for their volume-limited sample of Type IIP SN progenitors; beyond this distance many massive stars are too faint too detect, and the spatial resolution of {\it HST} causes problems with crowding and blending of point sources.

We consider only data from the three most recent wide-field optical cameras on {\it HST}, namely the Advanced Camera for Surveys / Wide Field Channel (ACS/WFC), the Wide Field and Planetary Camera 2 (WFPC2) and the Wide Field Camera 3 / Ultraviolet-Visible Channel (WFC3/UVIS). The characteristics of each of these instruments are listed in Table \ref{tab:inst}. Images from the High Resolution Channel (HRC) of ACS were not used, as the field of view of the camera (29\arcsec$\times$25\arcsec) is significantly smaller.

\begin{table}
 \begin{center}
 \caption{Field of view and spatial resolution of the {\it HST} instruments used in this work.}
 \label{tab:inst}
 \begin{tabular}{lcc}
  Instrument	& Field of view	& Pixel scale	\\
  \hline
  WFPC2		& 150\arcsec$\times$150\arcsec		& 0.10\arcsec	\\	
  ACS/WFC	& 202\arcsec$\times$202\arcsec		& 0.05\arcsec	\\	
  WFC3/UVIS	& 162\arcsec$\times$162\arcsec		& 0.04\arcsec	\\	
 \end{tabular}
 
 \end{center}
\end{table}

\begin{figure}
\includegraphics[angle=270,scale=0.70]{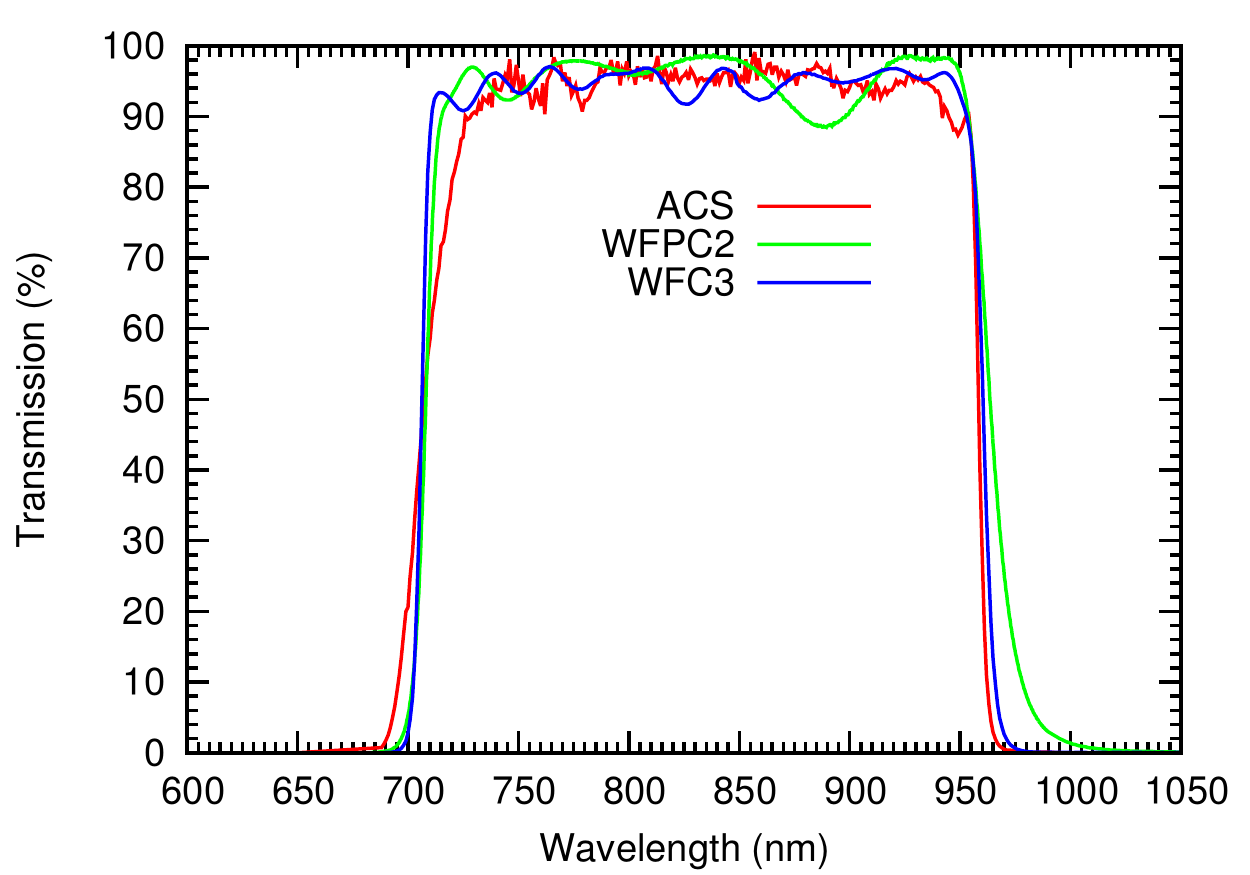}
\caption{Normalized transmission functions for the {\it F814W} filters of WFPC2, ACS and WFC3 onboard {\it HST}.}
\label{fig:filters}
\end{figure}

As the targets we are searching for are cool RSGs, the red {\it F814W} ($\sim I$-band) filter is most sensitive to them. Furthermore, {\it F814W} is among the more commonly used {\it HST} filters, and so data is available for a significant fraction of our sample of galaxies. The {\it F814W} filters of WFPC2, ACS and WFC3 are also quite similar, as can be seen in Fig. \ref{fig:filters}, allowing us to combine data from multiple instruments. As initial criteria for inclusion in our sample, we required a galaxy to have been observed on three separate epochs, with at least six months between the second and third epochs; and with at least some overlapping area common to all three epochs. We searched the Mikulski Archive for Space Telescopes (MAST)\footnote{http://archive.stsci.edu/} for any {\it HST} observations within a 5\arcmin~radius around the cataloged position of each of the galaxies in our sample, and which met these criteria. The observations for each galaxy were then examined using the visualisation tools in the Hubble Legacy Archive\footnote{http://hla.stsci.edu/} to determine whether their footprints overlapped. 

It is also necessary to have a minimum period between the first and second epochs, to ensure that short period variables and short-lived transients such as novae can be rejected. For most of our sample, we have at least 50 days separating the first and second epochs (as listed in Table \ref{tab:hla_data_1}), with only one galaxy (NGC 4639) having a shorter period. While ideally one would have several years separating the first and second epochs, we are limited in this work by the available archival data.

We found 76 galaxies within our input catalog which met our initial cuts, and had some overlap between the three epochs. For each of these galaxies, NED was queried for any non-kinematic measurements of the distance. Kinematic distances (as used in our initial selection) are prone to error for nearby galaxies where peculiar motions will be significant compared to recessional velocities. In our analysis of candidates, we use Cepheid and SN Ia luminosity distances to the host galaxy.

We also adopt a minimum distance limit of 10 Mpc for our sample. In the case of galaxies closer than this, the field of view of images from {\it HST} cover only a small fraction of a typical galaxy. Furthermore, in a typical 600s exposure with {\it HST}+ACS, a source with magnitude $m_\mathrm{I}$=25 is detected at $\sim10\sigma$, implying that for galaxies within $\sim10$ Mpc, many sources fainter than $M_{\mathrm{F814W}}<-5.2$ will be detected, which are too faint to be massive red supergiants. To avoid galaxies with high levels of internal extinction, only those with an inclination angle $i$ of $<$70\degree\ were included. As early type galaxies have minimal (if any) ongoing star formation, we do not expect to find massive red supergiants within them. Galaxies with a morphological type of S0 or earlier were hence excluded from our sample, as were dwarf galaxies (with an absolute magnitude $M_\mathrm{B}<-18$) which will contain fewer stars than massive spiral galaxies. After making these cuts, we were left with a sample of 15 galaxies, which are listed in Table \ref{tab:gal_sample}, along with their basic properties. The imposed cuts (distance, inclination angle, galaxy type and galaxy mass) should not bias our results in any obvious fashion, but simply reduce false positives or increase our chance of finding a disappearing massive RSG.

\begin{table*}
\begin{center}
\begin{minipage}{150mm}
\caption{Properties of the final galaxy sample in this work. Two of the galaxies have recessional velocities which imply a distance of slightly closer than 10 Mpc. However, these should not bias our results because it will not affect the detection of massive stars.}
\begin{tabular}{lcccccc}
\hline
Identifier		& Type	& $i$	 (deg) 	& $v$ (\kms) &  Distance\footnote{Distances are derived from the recessional velocity from NED, after correction for local motion. When analysing our candidates in Sect. \protect\ref{sect:candidates} we use Cepheid and Type Ia SN luminosity distances which are discussed in the candidate section.} (Mpc) & $M_\mathrm{B}$ (mag)	& $A_\mathrm{F814W}$ \\
\hline
M66  	    	& SABb	& 68  		& 788	 & 6.5 (0.5) & -21.3   		&   0.05 \\  
NGC3368		& SABa  	& 51  		& 941	 & 11.8 (0.8) & -21.0	  	&   0.04 \\ 			
NGC1566		& SABb  	& 48  		& 1221	 & 19.4 (1.4) & -21.3   		&   0.01  \\
NGC2841		& Sb        	& 65  		& 831	& 12.3 (0.9) & -21.0   		&   0.02 \\ 
NGC4496A   	& SBd     	& 39			& 1780	& 13.6 (1.0) & -20.2   		&   0.04  \\
NGC4321		& SABb    & 23			& 1677	& 14.1 (1.0) & -22.1   		&   0.04  \\
NGC4414		& Sc      	& 57  		& 911	& 9.03 (0.6) & -20.2   		&   0.03  \\
NGC3982		& SABb    	& 30			& 1354 	& 21.8 (1.5) & -19.9   		&   0.02  \\
NGC4038		& SBm     	& 52  		& 1559	& 23.3 (1.6) & -21.5   		&   0.07  \\
NGC4639		& Sbc     	& 51  		& 1088	& 13.9 (1.0) & -19.2	   	&   0.04  \\
NGC4651		& Sc      	& 50  		& 913	& 14.0 (1.0) & -19.7   		&   0.04  \\
NGC3370		& Sc      	& 56  		& 1357	& 23.3 (1.6) & -19.8   		&   0.05  \\
NGC3021		& Sbc     	& 56			& 1672	& 26.5 (1.9) & -19.9   		&   0.02  \\
NGC1058		& Sc      	& 58  		& 634	& 9.0 (0.6) & -18.7	   	&   0.10  \\
NGC3344		& Sbc     	& 19			& 698 	& 6.9 (0.5) & -19.8   		&   0.05  \\
\hline
\end{tabular}
\label{tab:gal_sample}
\end{minipage}
\end{center}
\end{table*}

We have also excluded from our sample galaxies cases where the data at one of the epochs consisted of a single, short ($<$120~s) exposure. In shallow images such as these, we cannot be certain that we do not detect an extant RSG simply due to low signal to noise. Only three galaxies were excluded on this basis (NGC 5806, NGC 6217 and NGC 6951). We also note that cosmic rays, which are typically identified using two consecutive exposures, may pose a problem where only a single image is available. While there were no such cases in our data, in principle cosmic rays can be reliably identified in single images using {\sc lacosmic} \citep{Van01}.

We note that at this point we have not imposed any constraints to restrict our sample to galaxies where a SN necessarily {\it would} have been observed. Within our distance limit, systematic surveys such as the Catalina Real Time Transient Survey \citep{Dra09}, the Palomar Transient Survey \citep{Law09}, the Lick Observatory SN Search \citep{Lea11}, PanSTARRS-1 \citep{Kai10} and ASAS-SN \citep{Sha14} ensure that most SNe are detected (although see \citealp{Pri12} for a counterexample). However, prior to 1998 most of these surveys were not in operation, and it is likely that some optically bright SNe within this volume were missed \citep{Sma09a}. Another reason why a SN could be missed is due to extinction, however in such a case the progenitor would almost certainly be obscured also. In order to reduce the possibility that a massive star in our sample exploded as an optically bright SN, but passed un-noticed, we could exclude any galaxies observed prior to 1998. A further requirement could be to only consider galaxies which are above +70$\degree$ or below -70$\degree$ in declination, as such circumpolar targets are visible for almost the entire year. This would reduce the chance that a SN exploded in a galaxy just as it went behind the Sun, and then was too faint to be detected at the start of the subsequent observing season\footnote{If one were to restrict the sample to galaxies at high or low equatorial latitudes, then one would need to ensure that ground based surveys were also monitoring these regions.}. However, as these constraints would reduce our sample size still further we have not imposed them here, but rather discuss their consequences in Section \ref{sect:discussion}.

\subsection{Alignment and subtraction}

The data products used in our survey were the Hubble Legacy Archive (HLA) pipeline-drizzled mosaics. These have been resampled from the original data, and hence are sub-optimal for photometry. However, this should not affect difference imaging to any significant extent. In one case, the observations of NGC 1566 from 1995 Aug 30, no mosaic was available in the HLA archive; here we created our own mosaic using the {\sc iraf}\footnote{{\sc iraf} is distributed by the National Optical Astronomy Observatory, which is operated by the Association of Universities for Research in Astronomy (AURA) under cooperative agreement with the National Science Foundation.} {\sc stsdas} package.

For each galaxy, before performing difference imaging, we registered all frames to a common pixel scale, defined to be that of the 
image taken at the first epoch. To achieve this, approximately 30 point sources which were present in all three images 
were identified. As far as possible, these sources were distributed evenly across the entire 
region of overlap. We avoided using foreground stars in our alignment, as these can show appreciable proper
motion in {\it HST} data taken over timescales of $\sim$years. The centroids of the point sources were measured in {\sc iraf}, 
and matched coordinate lists were used with {\sc iraf geomap} to derive a geometric transformation between the first and second, and 
first and third epochs. The ``general'' transform was used in this process, which accounts for shifts, rotation, scaling in x and y, and a polynomial term which accounts for any residual distortion in the images. After deriving a transformation, the images from the second and third epochs were then transformed into the pixel coordinate system of the image from the first epoch using {\sc iraf geotran}. The images were then checked visually to ensure that there was no obvious misalignment.

The different point spread functions (PSFs) of WFPC2, ACS and WFC3, in conjunction with the effects of resampling when aligning images, meant that it was not possible to directly subtract pairs of images from each other. Instead, one of the images was first convolved with a kernel so that the PSFs of both images matched, before the images were scaled to a common flux level and subtracted from each other. This technique of ``Optimal Image Subtraction'' \citep{Ala98} is commonly used in transient surveys to detect SNe, and has been used with {\it HST} images to confirm the disappearance of progenitor candidates \citep{Mau14}. We used the ``{\sc High Order Transformation of Point Spread Function and Template Subtraction}'' ({\sc hotpants}) package\footnote{http://www.astro.washington.edu/users/becker/v2.0/hotpants
.html} to perform image subtraction between the first and third epochs, and second and third epoch images for each galaxy in our sample. An example of this process is shown in Fig. \ref{fig:sub}.

\subsection{Source detection and candidate identification}

To identify sources in the difference images, we used the {\sc sextractor} package \citep{Ber96}. For a source to be detected in the difference images, we require four adjacent pixels to be at least 3$\sigma$ above the measured background. For an optically dark SN which is present in the first two epochs, and gone in the third epoch, there should be a positive source in both subtracted images (i.e. Epoch 1 - Epoch 3, and Epoch 2 - Epoch 3). We hence crossmatched the list of detected sources in each difference image, keeping only those which were present in both subtracted images, and coincident to $<$4 pixels. Our detection limit varies depending on the exposure time of the images, and on the local background flux at a given position; however, our candidates have {\it F555W} magnitudes of $\lesssim$23 -- 26.5 mag, and this is indicative of the depth to which we are sensitive.

\begin{figure*}
\centering
	\begin{minipage}[]{0.49\linewidth}
		\includegraphics[width=1\linewidth]{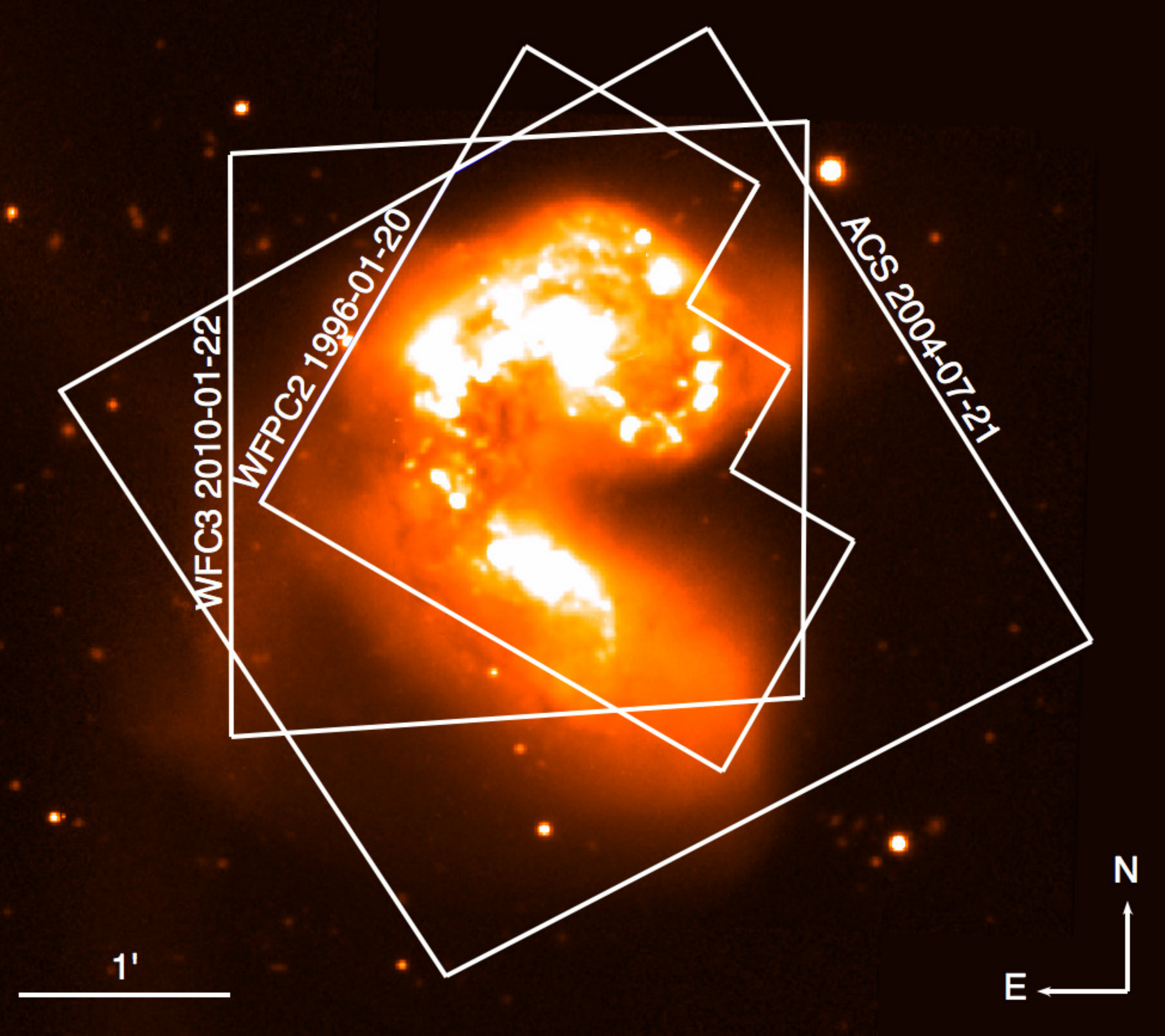}	
	\end{minipage}
	\begin{minipage}[]{0.49\linewidth}
		\includegraphics[width=1\linewidth]{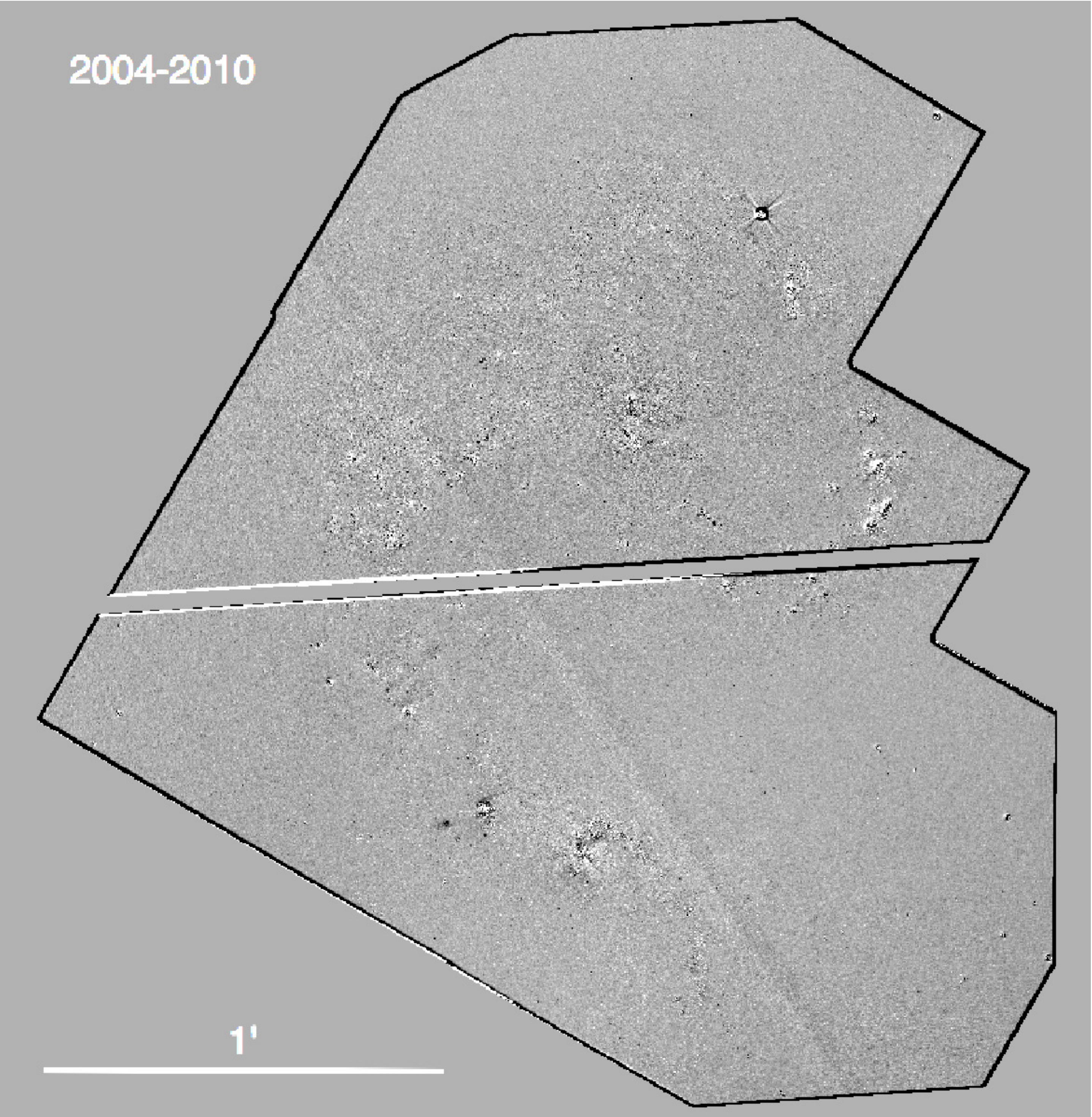}	
		\includegraphics[width=1\linewidth]{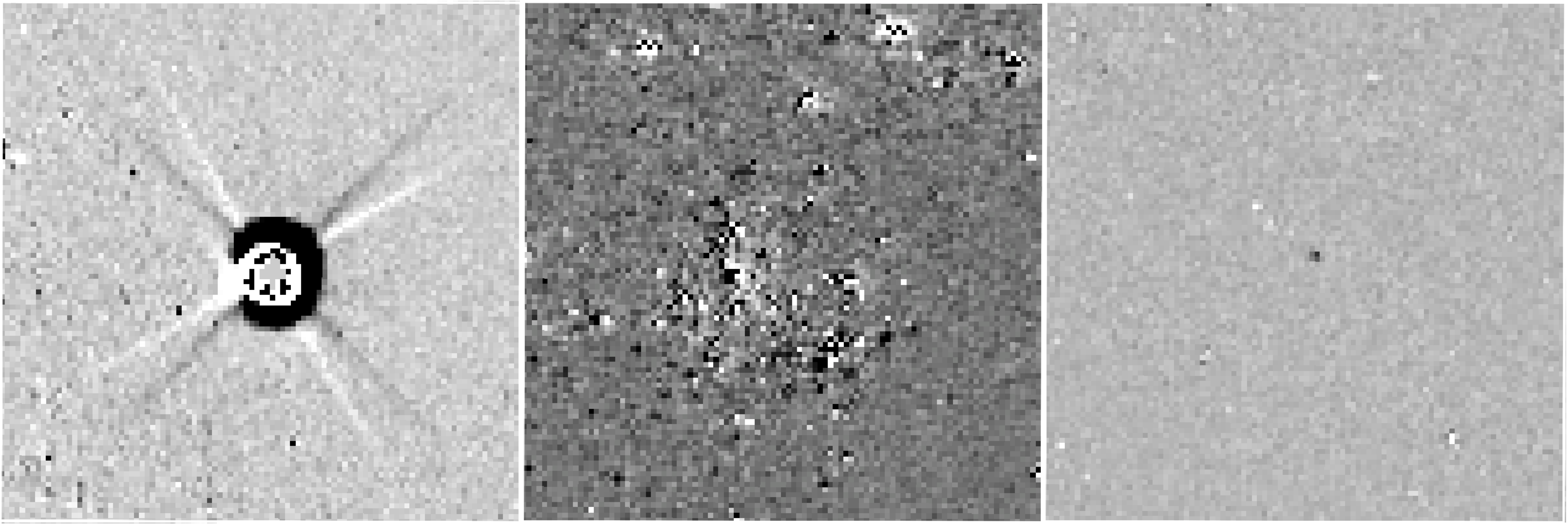}
	\end{minipage}
\caption[]{{\it Left}: The field of view of each of the {\it HST} images for  NGC 4038/4039 (the Antennae Galaxies), overlaid on an {\it R}-band image from the CTIO 0.9m telescope (via NED).
{\it Right}: The results of the image subtraction process for the 2004 and 2010 images of NGC 4038/4039. The upper panel shows the difference image for the entire galaxy pair, while the lower panels show a zoom-in on (from left to right) a bad subtraction from a saturated star, a bad subtraction of a dense knot in a spiral arm, and a good point source in the difference image.}
\label{fig:sub}
\end{figure*}

The source detection and crossmatch process typically yielded a list of 50--75 candidates per galaxy. The vast majority of these candidates are artefacts resulting from the subtraction process, such as residuals surrounding bright or saturated stars, or residuals in regions of complex background structure. Examples of such artefacts are shown in Fig. \ref{fig:sub}. To screen these, we performed a visual inspection of all candidates, removing bad subtractions, and only keeping those sources where there was no obvious source visible in the final epoch. At the end of this process we obtained a sample of six candidates, which are discussed in detail in Sect. \ref{sect:candidates}.

\subsection{Candidate photometry}

For each candidate, we searched the MAST archive for {\it all} WFPC2, ACS and WFC3 images covering the location of the candidate. These images were used in preference to the Hubble Legacy Archive products which were used to perform difference imaging and identify candidates, as the flux in these images has not been resampled. The reduced {\sc c0f} files (for WFPC2) and {\sc flt} or {\sc flc} files (for ACS and WFC3) were downloaded from the MAST archive.  

For the WFPC2 data, PSF-fitting photometry was performed with the {\sc hstphot} package \citep{Dol00a}. {\sc hstphot} has been specifically designed to perform optimally on undersampled WFPC2 images, and includes zeropoints, charge transfer efficiency (CTE) corrections and aperture corrections as per \cite{Dol00b}. Before performing photometry, the WFPC2 images were pre-processed using the ancillary routines distributed with {\sc hstphot}. This pre-processing included masking pixels which were flagged as bad in the data quality frame, combining multiple images taken at the same position to reject cosmic rays, masking hot pixels, and finally measuring the sky background for each pixel by taking the mean of the surrounding pixels. PSF fitting photometry was then performed on the processed images using pre-computed PSFs appropriate to each filter. The most up-to-date CTE corrections were applied to the measured photometry for each candidate, along with aperture corrections which were measured from bright stars in the images. The magnitudes for each candidate are listed in the VEGAMAG system

For the ACS and WFC3 images, we used the {\sc dolphot} package to perform photometry. {\sc dolphot} is a modified version of {\sc hstphot}, and includes libraries of PSFs, zeropoints and CTE corrections for each filter and instrument. Both the ACS and WFC3 images were pre-processed by masking bad pixels using the data quality frame for each image, and multiplying the image by a pixel area map to account for the non-uniform size of pixels due to geometric distortions across the instrument field of view. The local sky background was then measured for each pixel in the image, based on the surrounding pixels. Library PSFs were fitted to sources in each frame, and their magnitudes measured. For some of the ACS and WFC3 data, {\sc flc} files were available in the MAST archive; these have been corrected for CTE losses in the images and so no CTE correction was applied to measured magnitudes. For {\sc flt} files, CTE corrections were applied to the magnitude of each measure source. Aperture corrections were applied in all cases.

\begin{figure*}
\centering
	\includegraphics[width=0.33\linewidth,height=0.33\linewidth]{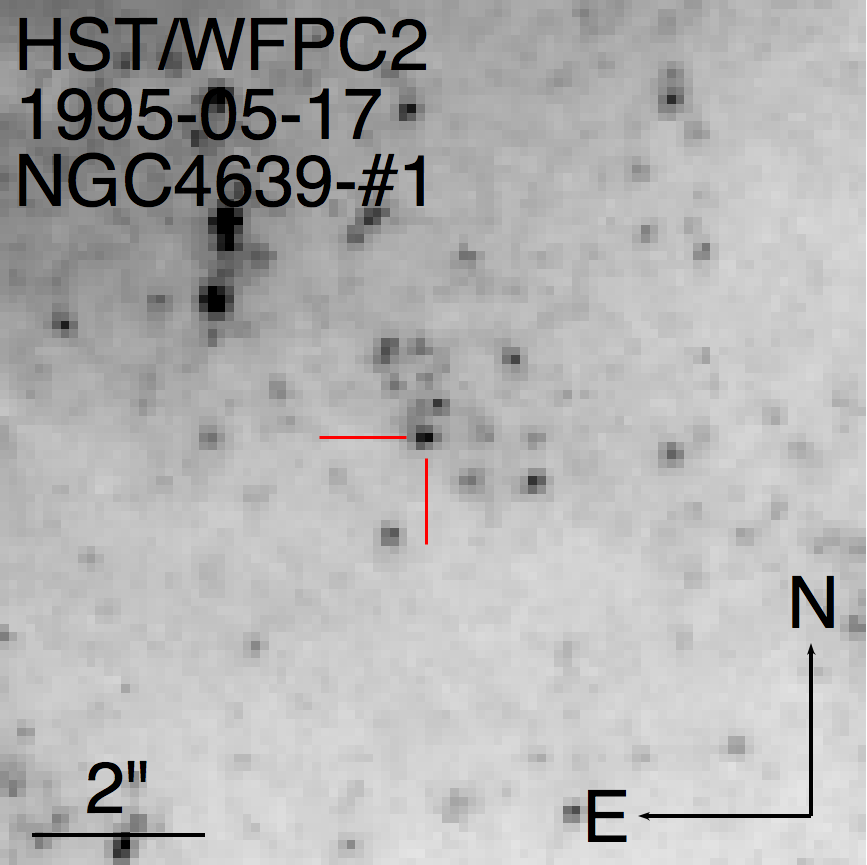}
	\includegraphics[width=0.33\linewidth,height=0.33\linewidth]{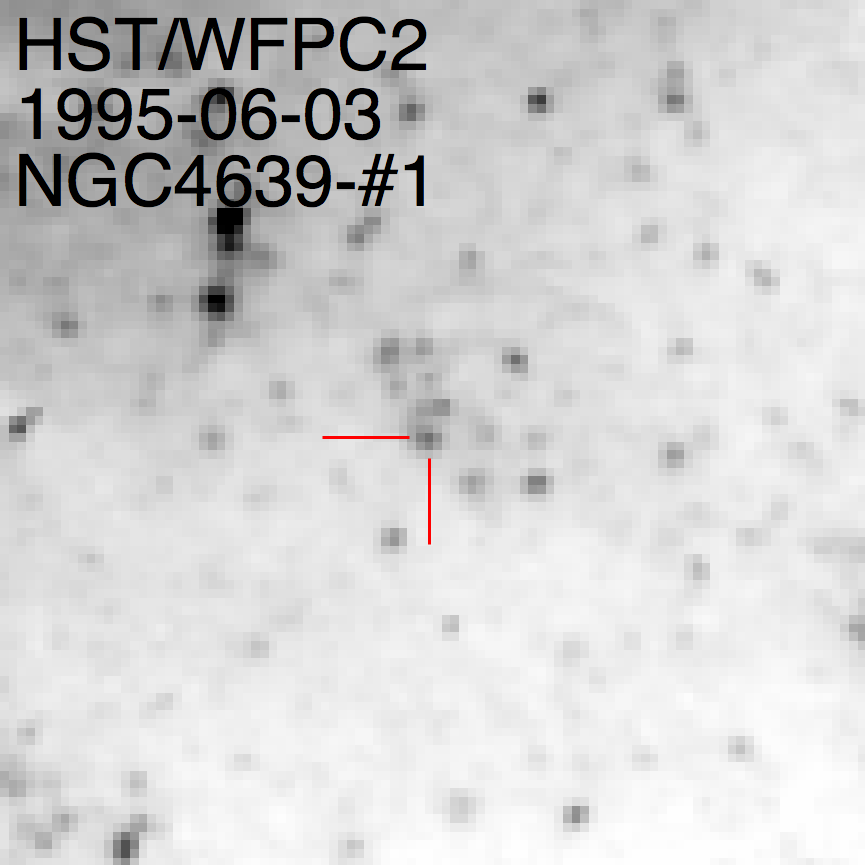}	
	\includegraphics[width=0.33\linewidth,height=0.33\linewidth]{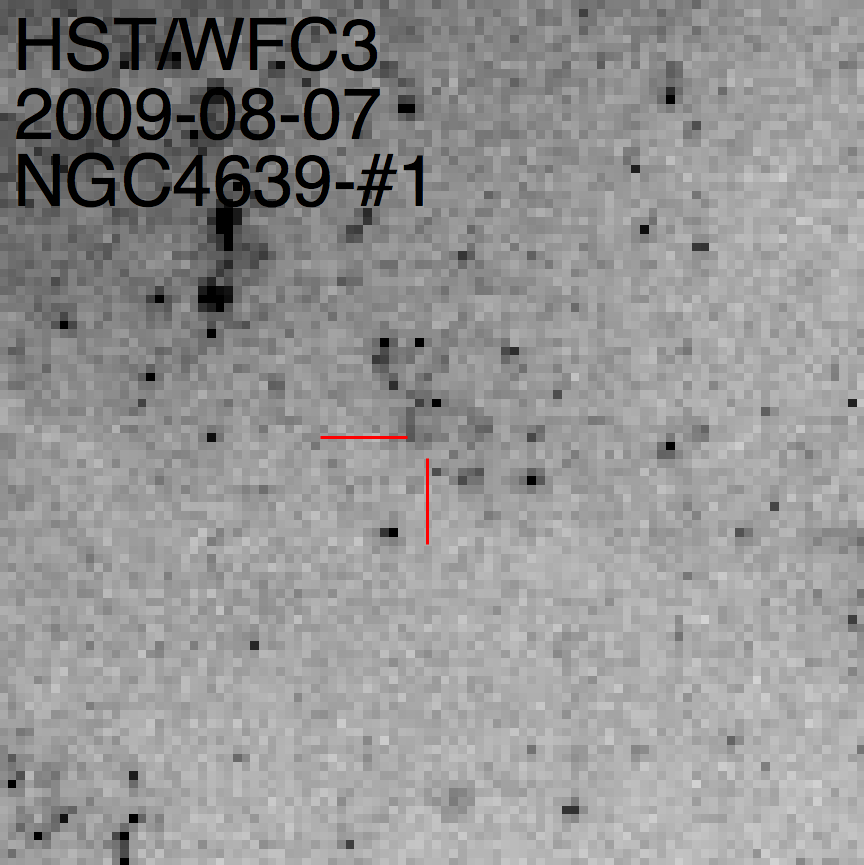}
\caption[]{{\it HST} {\it F814W} image cutouts centred on the position of NGC4639-CANDIDATE-1, the location of which is indicated with tick marks. All three panels cover a region 100$\times$100 pixels in size, with a common orientation and scale as indicated. The panels show from left to right the first, second and third epoch images used to search for candidates; the second and third epoch images have been aligned and resampled to match the first epoch as described in the text.}
\label{fig:NGC4639_1_stamps}
\end{figure*}

\begin{figure}
\centering
	\includegraphics[width=0.9\linewidth]{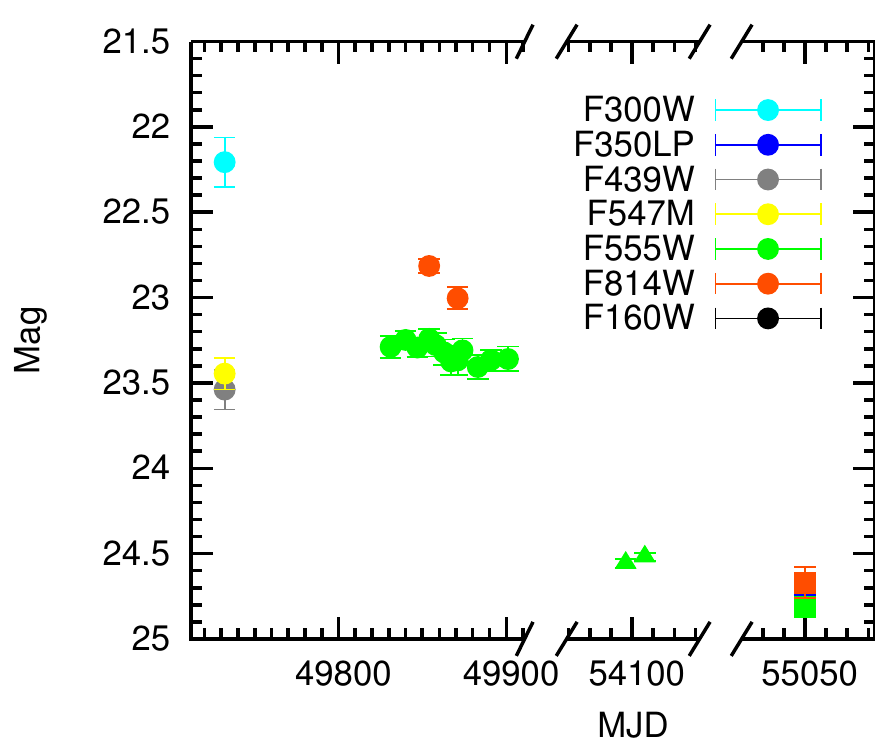}	
\caption[]{The lightcurve of NGC4639-CANDIDATE-1. Circles denote WFPC2 observations, triangles are ACS observations and squares are WFC3 observations.}
\label{fig:NGC4639-CANDIDATE-1_LC}
\end{figure}

\section{Candidates}
\label{sect:candidates}

In the following section, we present the analysis of each of the candidates identified.

\subsection{NGC4639-CANDIDATE-1}

The distance to NGC 4639 has been determined using Cepheids, mostly recently by \cite{Rie11} who derived a distance modulus of $\upmu$=31.67$\pm$0.08. This is consistent with the value of $\upmu$=31.71$\pm$0.08 found by \cite{Fre01}, although slightly closer than that obtained from the Type Ia SN 1990N \citep
[$\upmu=32.14\pm0.14$;][]{Rei05}. We adopt the \citeauthor{Rie11} distance in the following. The foreground extinction towards NGC 4639 is relatively low, $A_\mathrm{V}=0.071$~mag \citep{Sch11}. 

The coordinates of NGC4639-CANDIDATE-1 were found to be 12$^{\mathrm{h}}$42$^{\mathrm{m}}$52$^{\mathrm{s}}$.17 +13\degree14\arcmin54\arcsec.8, as measured from the HLA {\it F814W} mosaic. The lightcurve for NGC4639-CANDIDATE-1 is plotted in Fig. \ref{fig:NGC4639-CANDIDATE-1_LC}, while selected epochs of imaging are shown in Fig. \ref{fig:NGC4639_1_stamps}. The measured photometry for NGC4639-CANDIDATE-1 is listed in Table \ref{tab:NGC4639-CANDIDATE-1}. The {\it F814W}-band lightcurve shows a 1.5 mag drop between 1995 May / June, and 2009 August, when NGC4639-CANDIDATE-1 was detected by {\sc dolphot} at a much fainter magnitude. However, visual inspection of the 2009 images does not reveal an obvious point source, and so we cannot exclude the possibility that NGC4639-CANDIDATE-1 has disappeared completely in {\it F814W} in the last epoch, and that {\sc dolphot} is simply detecting some residual background structure at its location.

\begin{table}
\begin{center}
\caption{Photometry for NGC4639-CANDIDATE-1; associated errors are given in parentheses.}
\begin{tabular}{lcccccccr}
\hline
Date			& Instrument	& Filter	& Mag			\\
\hline
1995-01-14/15	& WFPC2		& F300W 	& 22.207 (0.145)  	\\
-			& -			& F439W 	& 23.539 (0.115)   	\\
-			& -			& F547M 	& 23.446 (0.093) 	\\
1995-04-24	& -			& F555W 	& 23.289 (0.065)	\\
1995-05-03	& -			& - 		& 23.248 (0.052)  	\\
1995-05-10	& -			& - 		& 23.293 (0.057) 	\\
1995-05-17	& -			& - 		& 23.242 (0.057)	\\
-			& -			& F814W 	& 22.814 (0.041) 	\\
1995-05-21	& -			& F555W	& 23.278 (0.068)	\\
1995-05-26	& -			& -		& 23.321 (0.076)	\\
1995-05-30	& -			& -		& 23.375 (0.076) 	\\
1995-06-03	& -			& -		& 23.369 (0.086) 	\\
-			& -			& F814W	& 23.003 (0.064)	\\
1995-06-06	& -			& F555W	& 23.309 (0.069)	\\
1995-06-15	& -			& -		& 23.406 (0.071)	\\
1995-06-23	& -			& -		& 23.367 (0.063)	\\
1995-07-03	& -			& -		& 23.359 (0.073)	\\
2006-12-28	& ACS/ WFC1	& F555W	& 24.556 (0.027)	\\
2007-01-06	& -			& -		& 24.520 (0.025)  	\\
2009-08-07	& WFC3/UVIS1	& F555W 	& 24.812 (0.042)	\\
-			& -			& F814W	& 24.671 (0.091)	\\
-			& -			& F350LP	& 24.780 (0.038)	 \\
\hline
\end{tabular}
\label{tab:NGC4639-CANDIDATE-1}
\end{center}
\end{table}

\begin{figure*}
\centering
	\includegraphics[width=0.33\linewidth,height=0.33\linewidth]{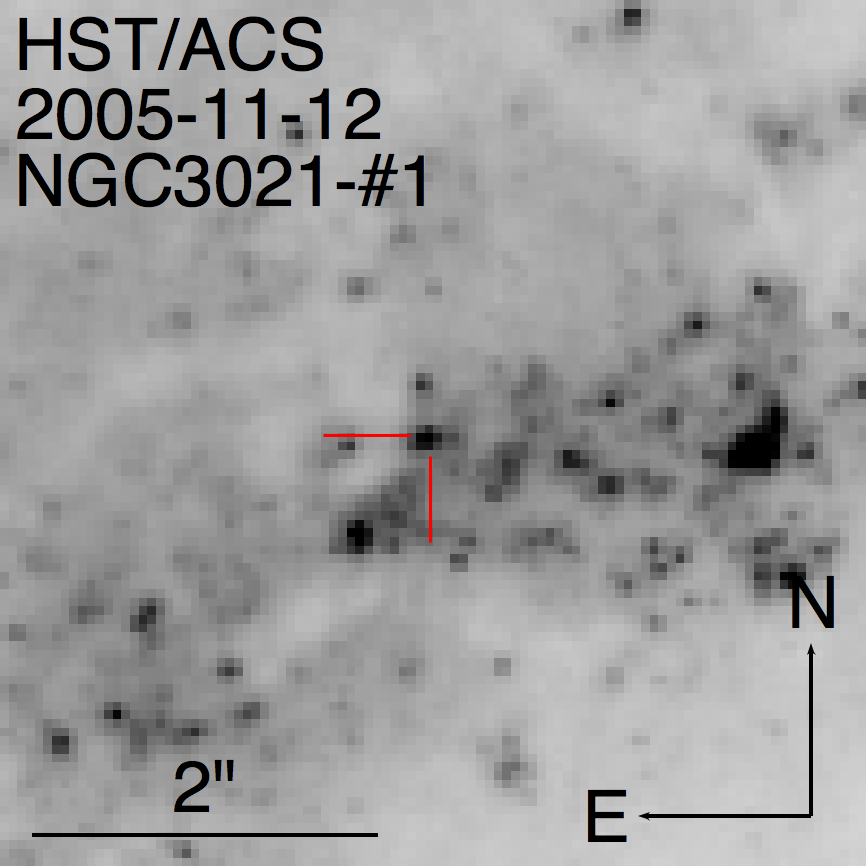}
	\includegraphics[width=0.33\linewidth,height=0.33\linewidth]{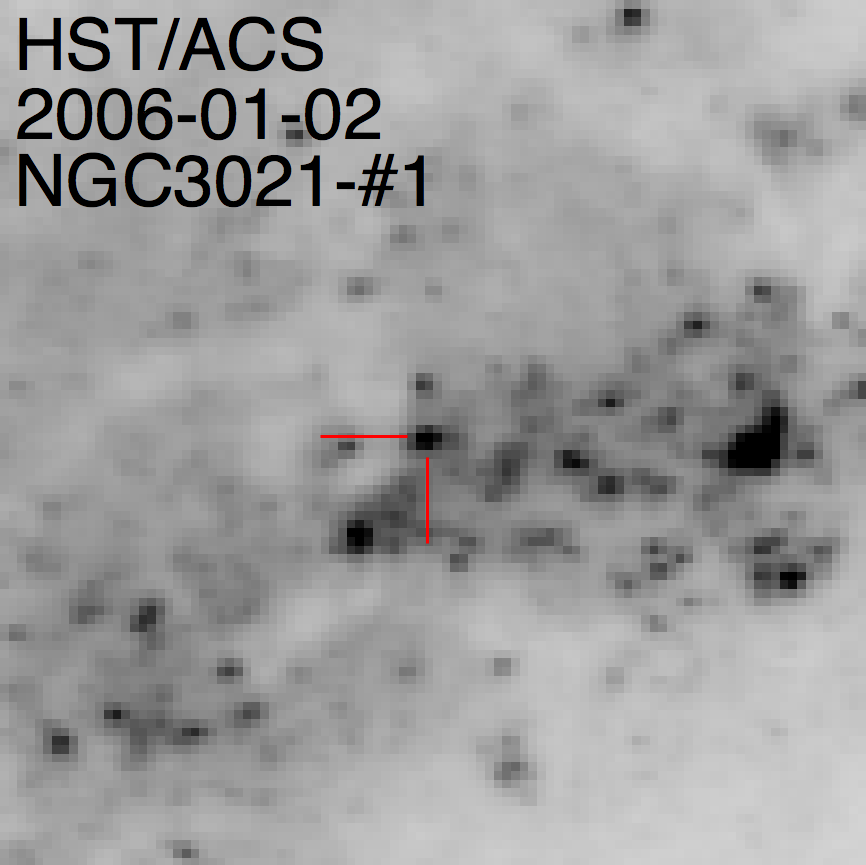}	
	\includegraphics[width=0.33\linewidth,height=0.33\linewidth]{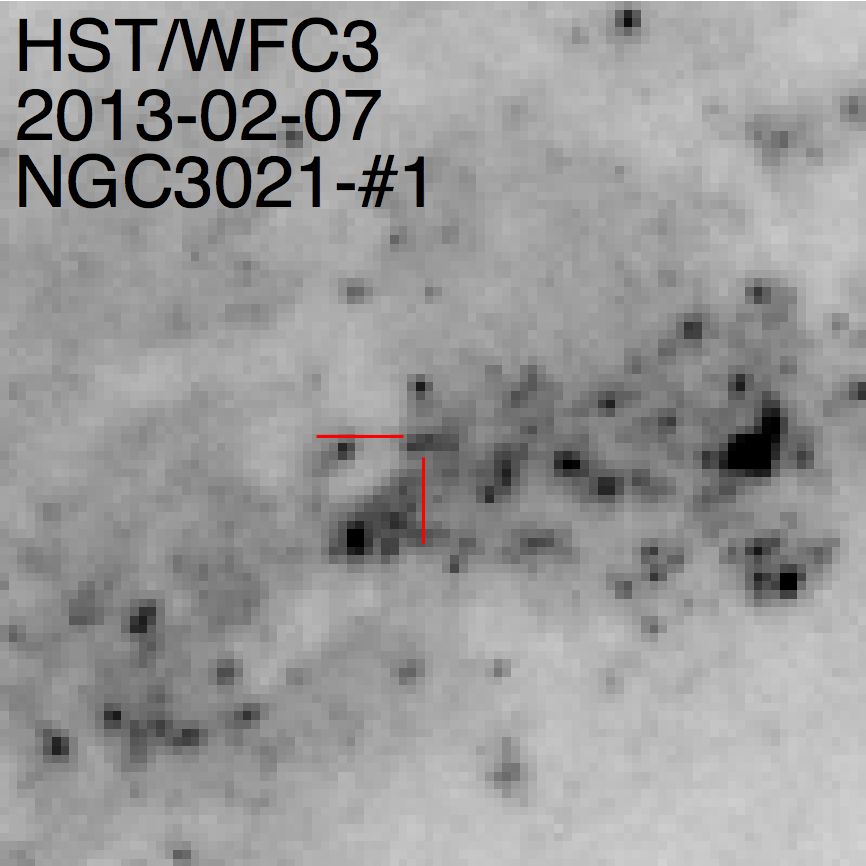}
\caption[]{{\it HST} {\it F814W} image cutouts centred on the position of NGC3021-CANDIDATE-1, similar to Fig. \ref{fig:NGC4639_1_stamps}}
\label{fig:NGC3021_1_stamps}
\end{figure*}

We calculate the average {\it F814W} and {\it F555W} magnitudes of NGC4639-CANDIDATE-1 over the period April - May 1995 to be $23.32\pm0.05$ and $22.91\pm0.13$ mag respectively. Using the adopted extinction and distance towards NGC4639, we derive absolute magnitudes of $M_{\mathrm{F555W}}=-8.42\pm0.09$ and $M_{\mathrm{F814W}}=-8.80\pm0.15$, and a colour {\it F555W}$-${\it F814W} = $0.38\pm0.14$ for NGC4639-CANDIDATE-1.

\begin{figure}
\centering
	\includegraphics[width=0.9\linewidth]{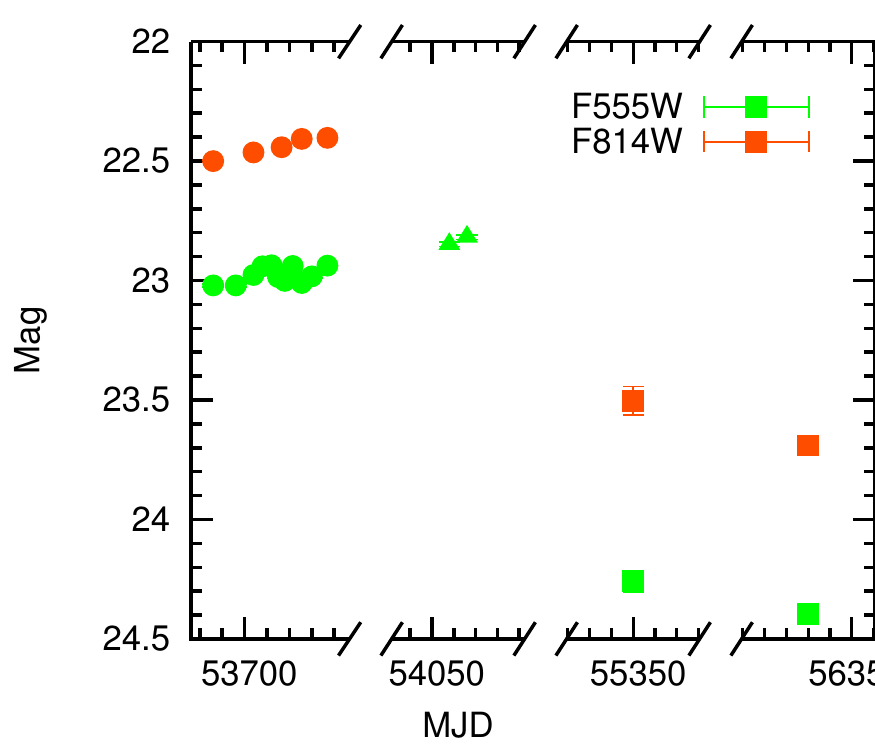}	
\caption[]{The lightcurve of NGC3021-CAND-1. Triangles denote ACS observations, squares are WFC3 observations. Minor tick marks are 10 d apart}
\label{fig:NGC3021-CAND-1_LC}
\end{figure}

The earliest set of images covering the site of NGC4639-CANDIDATE-1 were taken in 1995 using WFPC2, with the {\it F218W}, {\it F300W}, {\it F439W} and {\it F547M} filters. No source was detected in the {\it F218W}-filter image. In 1995, NGC4639-CANDIDATE-1 was brightest in {\it F300W}, suggesting that it was relatively hot, and perhaps in outburst at this point. Similarly, the robust detection of NGC4639-CANDIDATE-1 in the {\it F555W} filter in 2006 December and 2007 January, at approximately 1.2 mags fainter than in 1995, points towards a variable source, rather than one which has disappeared entirely by 2009.

\subsection{NGC3021-CANDIDATE-1}

We adopt the Cepheid derived distance modulus of $\upmu=32.27\pm0.08$ mag for NGC 3021 from \cite{Rie11}. The foreground extinction is taken to be $A_\mathrm{V}$=0.037.

Using the 2005 HLA {\it F814} mosaic, we measured coordinates of 09$^{\mathrm{h}}$50$^{\mathrm{m}}$55$^{\mathrm{s}}$.39 +33\degree33\arcmin14\arcsec.5 for NGC3021-CANDIDATE-1. The region of the candidate is show in Fig. \ref{fig:NGC3021_1_stamps}, while the photometry is listed in Table \ref{tab:phot_NGC3021-CAND-1} and plotted as a lightcurve in Fig. \ref{fig:NGC3021-CAND-1_LC}.

\begin{table}
\begin{center}
\caption{Photometry for NGC3021-CANDIDATE-1}
\begin{tabular}{lcccccccr}
\hline
Date			& Instrument	& Filter	& Mag				\\
\hline		
2005-11-12	& ACS/WFC2	& F555W	& 23.020 (0.007)	\\
-			& -			& F814W	& 22.500 (0.006)	\\
2005-11-22	& -			& F555W	& 23.020 (0.007)	\\
2005-11-30	& -			& F555W	& 22.977 (0.006)	\\
-			& -			& F814W	& 22.464 (0.006)	\\
2005-12-04	& -			& F555W	& 22.940 (0.007)	\\
2005-12-08	&-			& F555W	& 22.936 (0.006)	\\
2005-12-11	& -			& F555W	& 22.985 (0.006)	\\
2005-12-12/13	& -			& F555W	& 22.977 (0.006)	\\
-			& -			& F814W	& 22.442 (0.006)	\\
2005-12-14	& -			& F555W	& 23.002 (0.007)	\\
2005-12-17/18	& -			& F555W	& 22.939 (0.006)	\\
2005-12-21/22	& -			& F555W	& 23.010 (0.007)	\\
-			& -			& F814W	& 22.407 (0.006)	\\
2005-12-26	& -			& F555W	& 22.982 (0.007)	\\
2006-01-02	& -			& F555W	& 22.937 (0.006)	\\
-			& -			& F814W	& 22.402 (0.006)	\\
2006-11-19	& -			& F555W	& 22.849 (0.009)	\\
2006-11-27	& -			& F555W	& 22.819 (0.009)	\\
2010-06-03	& WFC3/UVIS2	& F555W	& 24.258 (0.045)	\\	
-			& -			& F814W	& 23.503 (0.059)	\\
2013-02-07	& WFC3/UVIS1	& F555W	& 24.395 (0.028) 	\\
-			& -			& F814W	& 23.690 (0.041)  	\\
\hline
\end{tabular}
\label{tab:phot_NGC3021-CAND-1}
\end{center}
\end{table}

From 2005 Nov 12, until 2006 Jan 2, NGC3021-CANDIDATE-1 had average magnitudes of {\it F555W}=$22.98\pm0.03$ and {\it F814W}=$22.44\pm0.04$. A year later, over the period 2006 November 19-27, the source was only 0.15 mag fainter, at {\it F555W} = $22.83\pm0.02$ mag. In 2005/2006, the source had an absolute magnitudes of $-9.33\pm0.09$ and $-9.85\pm0.09$ mag in {\it F555W} and {\it F814W} respectively, given the distance modulus and foreground extinction towards NGC 3021. The {\it F555W}$-${\it F815W} colour of NGC3021-CANDIDATE-1 was $0.37\pm0.04$ mag. Some 3.5 years later, NGC3021-CANDIDATE-1 appears to be $\sim$1.5 mag fainter in {\it F555W} and $\sim$1 mag fainter in {\it F814W}. A final set of observations in early 2013 show the source to be similarly faint.

\begin{figure}
\centering
	\includegraphics[width=0.9\linewidth]{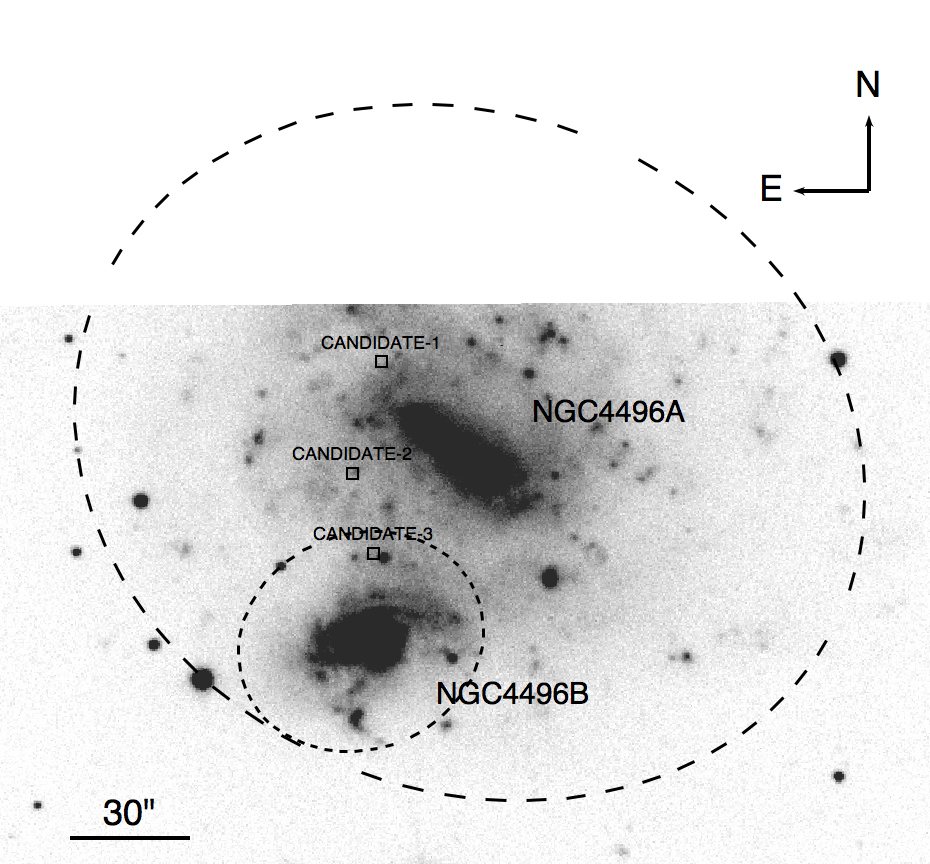}	
\caption[]{SDSS {\it r'} image of NGC4496A and NGC4496B. Each galaxy is indicated, the dashed ellipses represent the isophotal major and minor axes at a surface brightness level $B$=25 mag/arcsec$^{2}$). The positions of each of the candidates are marked.}
\label{fig:NGC4496AB}
\end{figure}

Whether the source detected by {\sc dolphot} at late times is related to NGC3021-CANDIDATE-1 is not immediately clear. From examination of the images (Fig. \ref{fig:NGC3021_1_stamps}), it appears that there is some extended background flux at the location of the candidate. There may be a contribution to this flux from the progenitor, or it may be comprised entirely of unrelated background sources. The region of the candidate shows several such partially resolved extended sources, so it would not be surprising to find a source in close proximity to one of these. Alternatively, as massive stars tend to be found in star forming complexes, this may be the host cluster of NGC3021-CANDIDATE-1. If this is the case, then NGC3021-CANDIDATE-1 clearly dominated the flux of the cluster and was its most massive inhabitant, which would be consistent with it being the next star to explode.

\subsection{NGC4496-CANDIDATE-1}

NGC4496-CANDIDATE-1 is associated with a galaxy pair, NGC 4496A and NGC 4496B, which are also designated VV76.  While NGC 4496A and NGC 4496B were originally thought to be a physically related pair of galaxies, they have significantly different recessional velocities of 1700 \kms\ and 4510 \kms\ respectively \citep{Fil88}. NGC 4496B, which has a higher recessional velocity has a smaller angular size, and is most likely behind NGC 4496A; while such coincident superpositions are rare, NGC 4496A and NGC 4496B lie toward the Virgo cluster which has a higher density of nearby galaxies.

\begin{table}
\begin{center}
\caption{Photometry for NGC4496-CANDIDATE-1}
\begin{tabular}{lcccccccr}
\hline
Date			& Instrument	& Filter	& Mag	\\
\hline	
1994-05-27	& WFPC2/WF4	& F555W	&  26.199 (0.137)  \\
1994-06-06	& -			& F555W	&  26.485 (0.266)  \\
-			& -			& F814W	&  23.850 (0.035)  \\
1994-06-11	&-			& F555W	&  26.251 (0.134)  \\
1994-06-16	&-			& F555W	&  26.097 (0.103)  \\
1994-06-18	&-			& F555W	&  26.091 (0.159)  \\
1994-06-21	&-			& F555W	&  26.132 (0.149)  \\
 -			&-			& F814W	&  23.773 (0.034)  \\
1994-06-22	&-			& F555W	&  25.966 (0.160)   \\
1994-06-26	&-			& F555W	&  26.128 (0.169)   \\
1994-06-29	&-			& F555W	&  26.128 (0.169)   \\
1994-07-01	&-			& F555W	&  26.152 (0.157)   \\
1994-07-03	&-			& F555W	&  26.260 (0.120)   \\
1994-07-16	&-			& F555W	&  26.239 (0.236)  \\
-			&-			& F814W	&  23.829 (0.043) \\
1994-07-20	&-			& F555W	&  26.258 (0.172)  \\
1994-07-26	&-			& F555W	&  26.014 (0.167)   \\
1994-08-02	&-			& F555W	&  26.053 (0.025)  \\
1994-08-06	&-			& F555W	&  26.355 (0.259)  \\
-			& -			& F814W	&  23.833 (0.043)\\
1994-08-07	&-			& F555W	&  26.289 (0.193)  \\ 
 2001-03-17	& WFPC2/WF2		& F814W	& $>$24.9    	 \\
\hline
\end{tabular}
\label{tab:NGC4496-SOURCE-1}
\end{center}
\end{table}

\begin{figure*}
\centering
	\includegraphics[width=0.33\linewidth,height=0.33\linewidth]{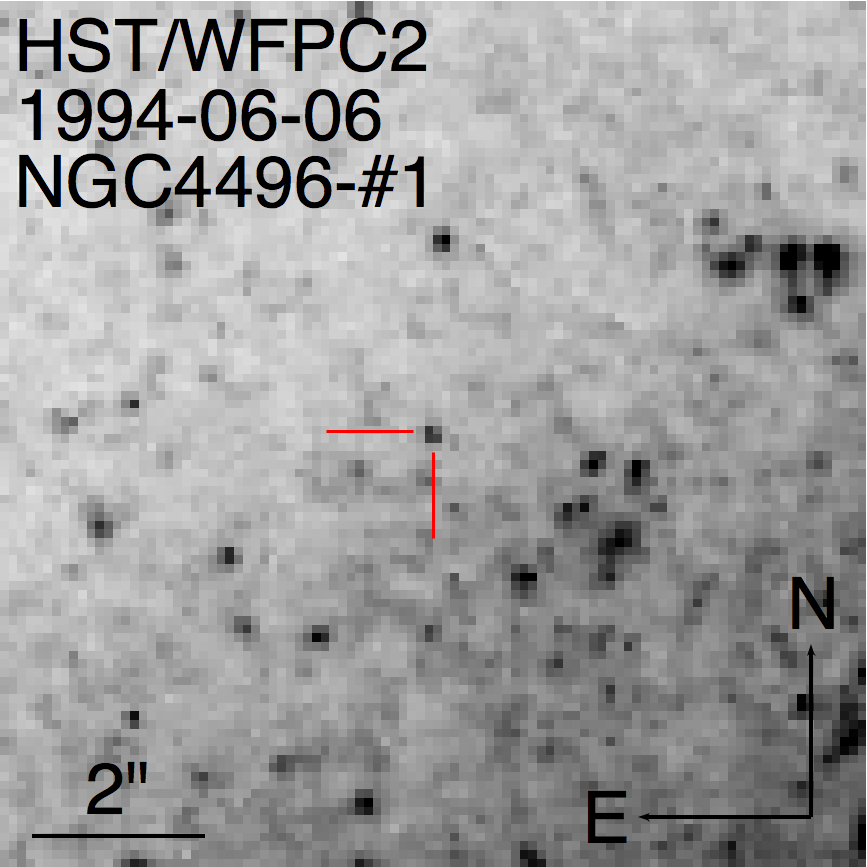}
	\includegraphics[width=0.33\linewidth,height=0.33\linewidth]{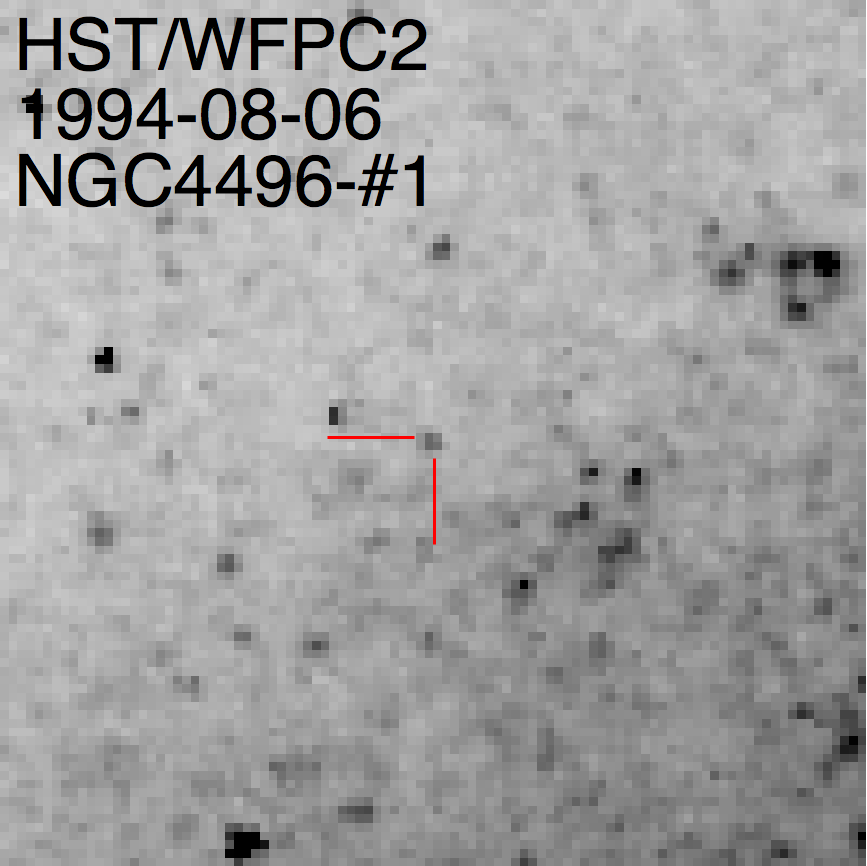}	
	\includegraphics[width=0.33\linewidth,height=0.33\linewidth]{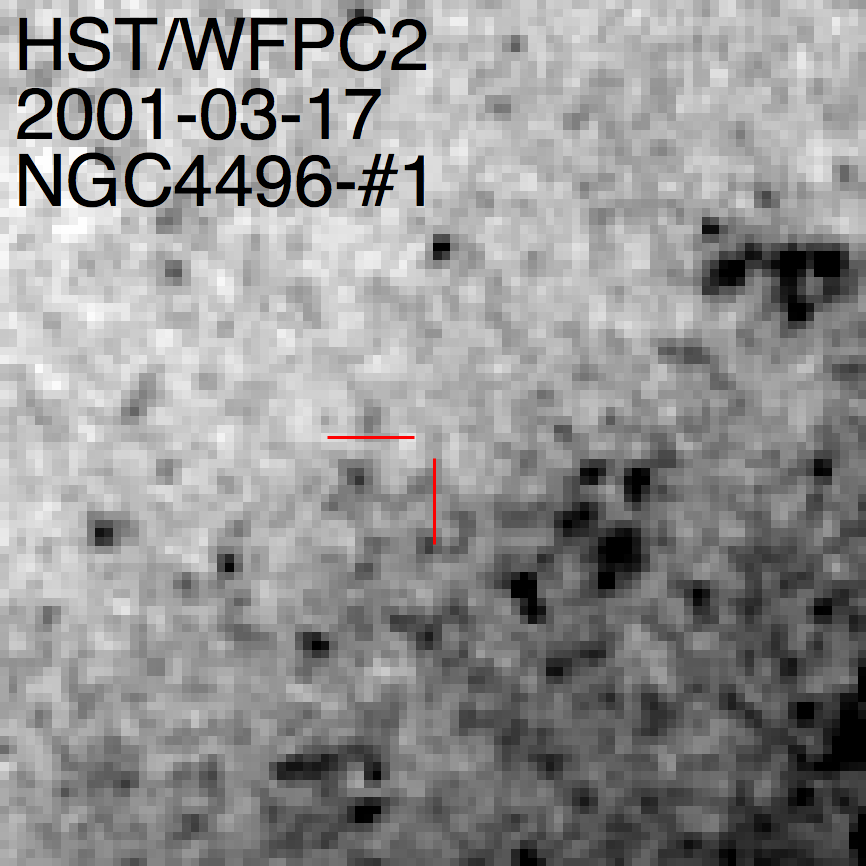}
\caption[]{{\it HST} {\it F814W} image cutouts centred on the position of NGC4496-CANDIDATE-1, similar to Fig. \ref{fig:NGC4639_1_stamps}.}
\label{fig:NGC4496_1_stamps}
\end{figure*}

We measured coordinates of 12$^{\mathrm{h}}$31$^{\mathrm{m}}$40$^{\mathrm{s}}$.57 +3\degree56\arcmin45\arcsec.1 for NGC4496-CANDIDATE-1 (Fig. \ref{fig:NGC4496_1_stamps}). The candidate lies within the isophotal radius of NGC 4496A at {\it B}~$=25$, and well outside the isophotal radius of NGC 4496B. NGC4496-CANDIDATE-1 also lies to the north of NGC 4496A, whereas NGC 4496B lies to the south. It hence appears almost certain that NGC4496-CANDIDATE-1 is associated with NGC 4496A.  The distance towards NGC 4496A has been measured both using Cepheids, and with the peak luminosity of the Type Ia SN 1960F. We adopt a Cepheid-based distance modulus of $30.86\pm0.03$ mag, corresponding to 14.86 Mpc \citep[from the final distance modulus with period cuts and correction for metallicity;][]{Fre01}. This is consistent with the luminosity distance from SN 1960F, 30.86 mag \citep{Rei05}. We adopt a foreground extinction of $A_\mathrm{V}=0.067$ mag towards both NGC 4496A and NGC 4496B.

\begin{figure}
\centering
	\includegraphics[width=0.9\linewidth]{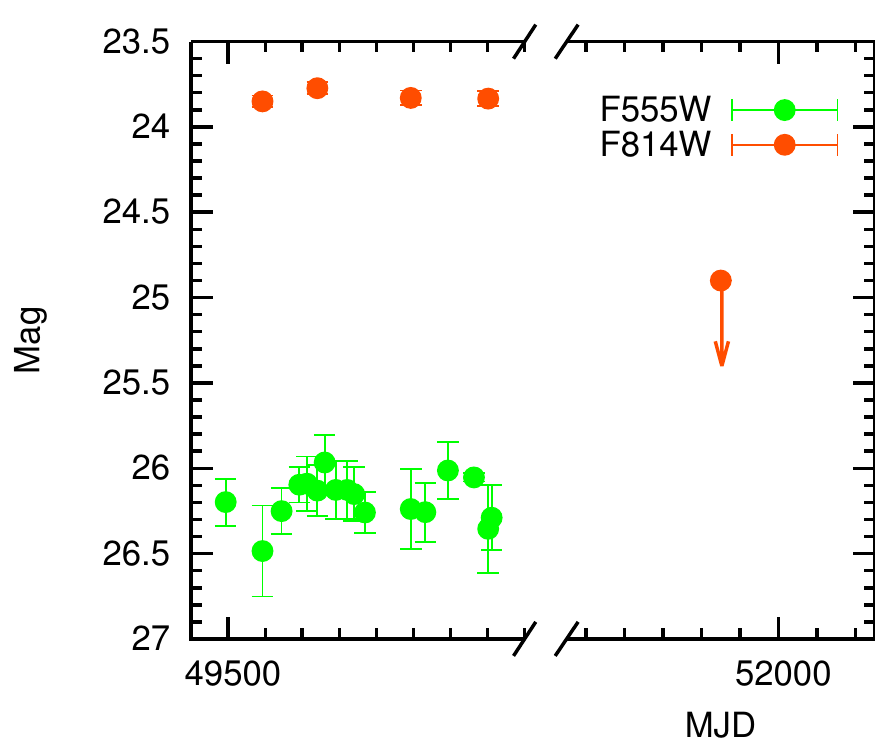}	
\caption[]{The lightcurve of NGC4496-CANDIDATE-1. Circles denote WFPC2 observations, minor tick marks are 10 d apart.}
\label{fig:NGC4496-CAND-1_LC}
\end{figure}

The measured magnitudes for NGC4496-CANDIDATE-1 are listed in Table \ref{tab:NGC4496-SOURCE-1}, and plotted as a lightcurve in Fig. \ref{fig:NGC4496-CAND-1_LC}. The source appears to have a constant magnitude in both {\it F555W} and {\it F814W} over the 71 days it was observed in mid-1994, with average magnitudes of $26.18\pm0.13$ and $23.82\pm0.03$ mag in {\it F555W} and {\it F814W} respectively. The quoted uncertainties on the average magnitude are the standard deviation of all measurements, which in both cases are comparable to the photometric uncertainty on individual measurements.

The absolute magnitude of NGC4496-CANDIDATE-1 in 1994 is relatively faint for a massive RSG. After correcting for distance and foreground extinction, the source has an absolute magnitude of {\it F555W}$=-4.75\pm0.13$ mag, and a {\it F555W}$-${\it F814W} colour of 2.33 mag. Such a magnitude is similar to that of the {\it faintest} RSGs, which have been claimed as sub-luminous Type IIP progenitors \citep{Fra11}. Nearly seven years later, in 2001 March, the source is no longer detected in {\it F814W} images. From surrounding sources, we set a 5$\sigma$ limit of {\it F814W}$>$24.9 mag on  NGC4496-CANDIDATE-1, implying it has faded by a least a magnitude.

\subsection{NGC4496-CANDIDATE-2}

\begin{figure*}
\centering
	\includegraphics[width=0.33\linewidth,height=0.33\linewidth]{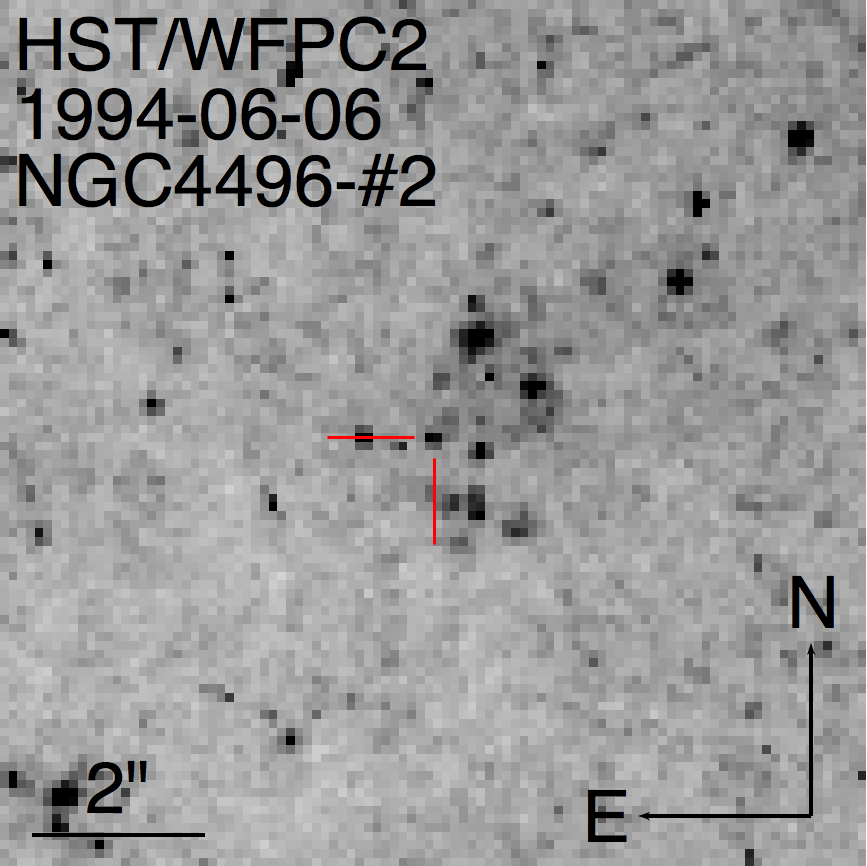}
	\includegraphics[width=0.33\linewidth,height=0.33\linewidth]{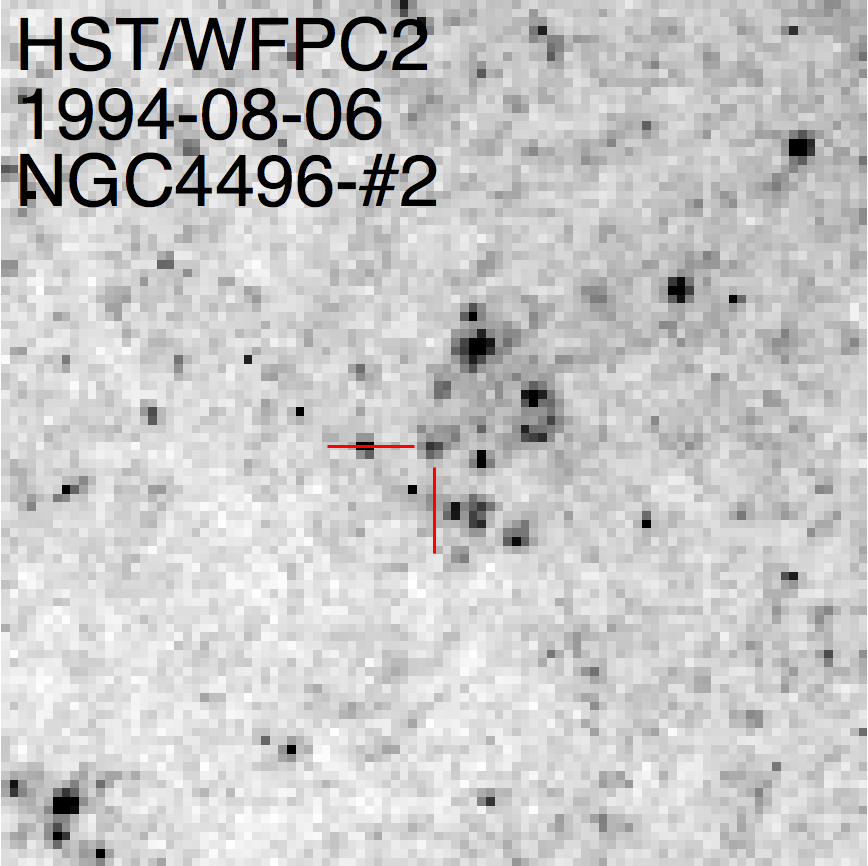}	
	\includegraphics[width=0.33\linewidth,height=0.33\linewidth]{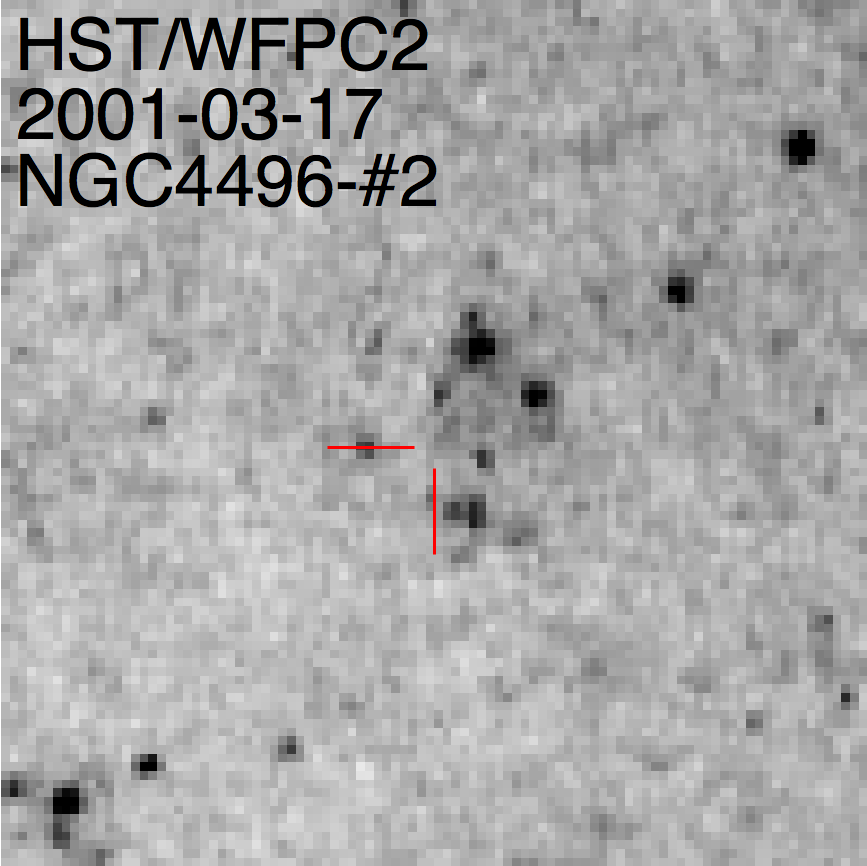}
\caption[]{{\it HST} {\it F814W} image cutouts centred on the position of NGC4496-CANDIDATE-2, similar to Fig. \ref{fig:NGC4639_1_stamps}.}
\label{fig:NGC4496_2_stamps}
\end{figure*}

\begin{figure}
\centering
	\includegraphics[width=0.9\linewidth]{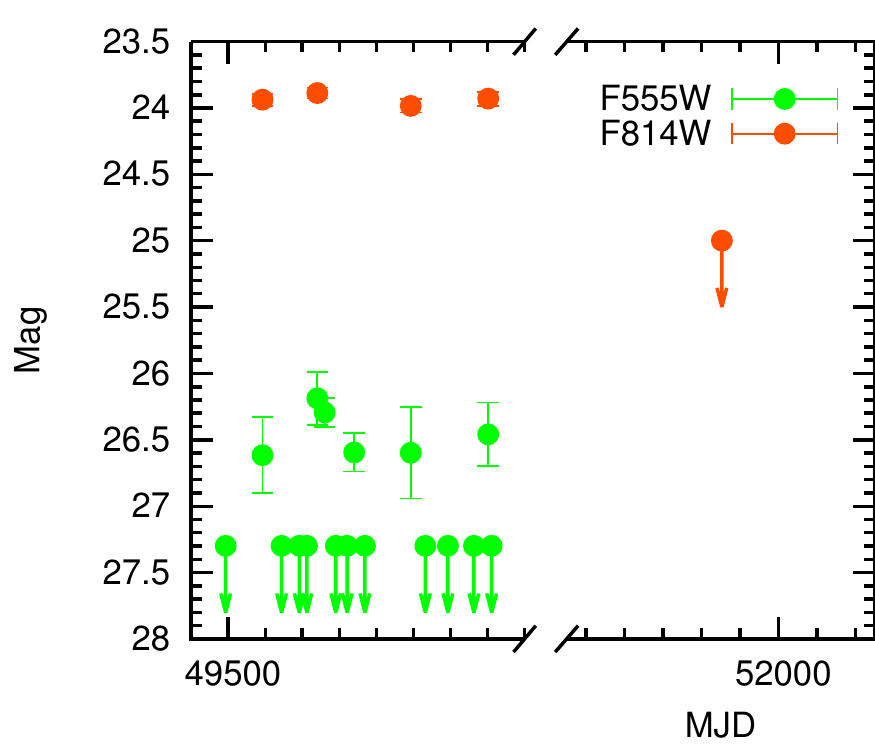}	
\caption[]{The lightcurve of NGC4496-CANDIDATE-2. Circles denote WFPC2 observations, minor tick marks are 10 d apart.}
\label{fig:NGC4496-CAND-2_LC}
\end{figure}

NGC4496-CANDIDATE-2 (Fig. \ref{fig:NGC4496_2_stamps}) was also found in the NGC 4496A/B pair. Similarly to NGC4496-CANDIDATE-1, NGC4496-CANDIDATE-2 lies closer to NGC4496A, is within its isophotal radius, and is outside the  radius of NGC 4496B. We adopt the same distance modus and foreground extinction ($\upmu=30.86\pm0.03$, and $A_\mathrm{V}=0.067$ mag) as used for NGC4496-CANDIDATE-1. We show the lightcurve of NGC4496-CANDIDATE-2 in Fig. \ref{fig:NGC4496-CAND-2_LC}, and report the magnitude of the source in each frame in Table \ref{tab:NGC4496-SOURCE-2}. The coordinates of NGC4496-CANDIDATE-2 are 12$^{\mathrm{h}}$31$^{\mathrm{m}}$41$^{\mathrm{s}}$.07 +3\degree56\arcmin17\arcsec.3.

NGC4496-CANDIDATE-2 is either not detected, or marginally detected in {\it F555W} in 1994. Over the first set of images, the average magnitude of NGC4496-CANDIDATE-2 is {\it F555W} = 26.46 $\pm$0.18, {\it F814W}~$= 23.93\pm0.04$ mag. This implies an absolute magnitude of  {\it F555W}~$= -4.47\pm0.18$ and  {\it F814W}~$= -6.97\pm0.05$ mag. We note that in many of the {\it F555W} images the source is not detected; this is not surprising, as the source is quite faint in this filter. The source is detected in all {\it F555W} images when there is contemporaneous {\it F814W} data, as we can use the latter to determine the position of the source in the former (effectively carrying out forced photometry at the position of the source in the {\it F555W} image). As in the case of NGC4496-CANDIDATE-1, if NGC4496-CANDIDATE-2 is a RSG, then it would be relatively faint, and hence low mass.

An alternative interpretation for NGC4496-CANDIDATE-2 is that it is in fact associated with NGC 4496B to the south. The distance to NGC 4496B based on its recessional velocity (after correction for infall on Virgo) is 70.1 Mpc, which gives a distance modulus of 34.23 mag. This would make the absolute magnitude of NGC4496-CANDIDATE-2 $-7.84$ mag in {\it F555W} and $-10.34$ mag in {\it F814W}. However, at this distance the implied galacto-centric radius of NGC4496-CANDIDATE-2 would be 14.6 kpc, which is quite large. On this basis, and on the fact that NGC4496-CANDIDATE-2 lies well outside the isophotal radius of NGC 4496B, we regard it as more likely to be associated with NGC 4496A.

\begin{table}
\begin{center}
\caption{Photometry for NGC4496-CANDIDATE-2}
\begin{tabular}{lcccccccr}
\hline
Date			& Instrument	& Filter	& Mag				\\
\hline	
1994-05-27	& WFPC2/PC	& F555W	& $<$27.3     \\
1994-06-06	&-			& F555W	& 26.616   (0.285)    \\
	-		& -			& F814W	& 23.938   (0.045)	\\
1994-06-11	&-			& F555W	& $<$27.3 \\
1994-06-16	&-			& F555W	& $<$27.3  \\
1994-06-18	&-			& F555W	& $<$27.3     \\
1994-06-21	&-			& F555W	& 26.187   (0.200)    \\
 	-		&-			& F814W	& 23.887   (0.041)  	\\
1994-06-22	&-			& F555W	& 26.293   (0.108)    \\
1994-06-26	&-			& F555W	& $<$27.3     \\
1994-06-29	&-			& F555W	& $<$27.3     \\
1994-07-01	&-			& F555W	& 26.595   (0.144)   \\
1994-07-03	&-			& F555W	& $<$27.3     \\
1994-07-16	&-			& F555W	& 26.598   (0.344)    \\
	-		&-			& F814W	& 23.983   (0.051)  	\\
1994-07-20	&-			& F555W	& $<$27.3     \\
1994-07-26	&-			& F555W	& $<$27.3     \\
1994-08-02	&-			& F555W	& $<$27.3     \\
1994-08-06	&-			& F555W	& 26.460   (0.241)    \\
	-		& -			& F814W	& 23.930   (0.053)  	\\
1994-08-07	&-			& F555W	& $<$27.3     \\
 2001-03-17	& WFPC2/WF4	& F814W	& $<$25.000    \\
\hline
\end{tabular}
\label{tab:NGC4496-SOURCE-2}
\end{center}
\end{table}

\subsection{NGC4496-CANDIDATE-3}

\begin{figure*}
\centering
	\includegraphics[width=0.33\linewidth,height=0.33\linewidth]{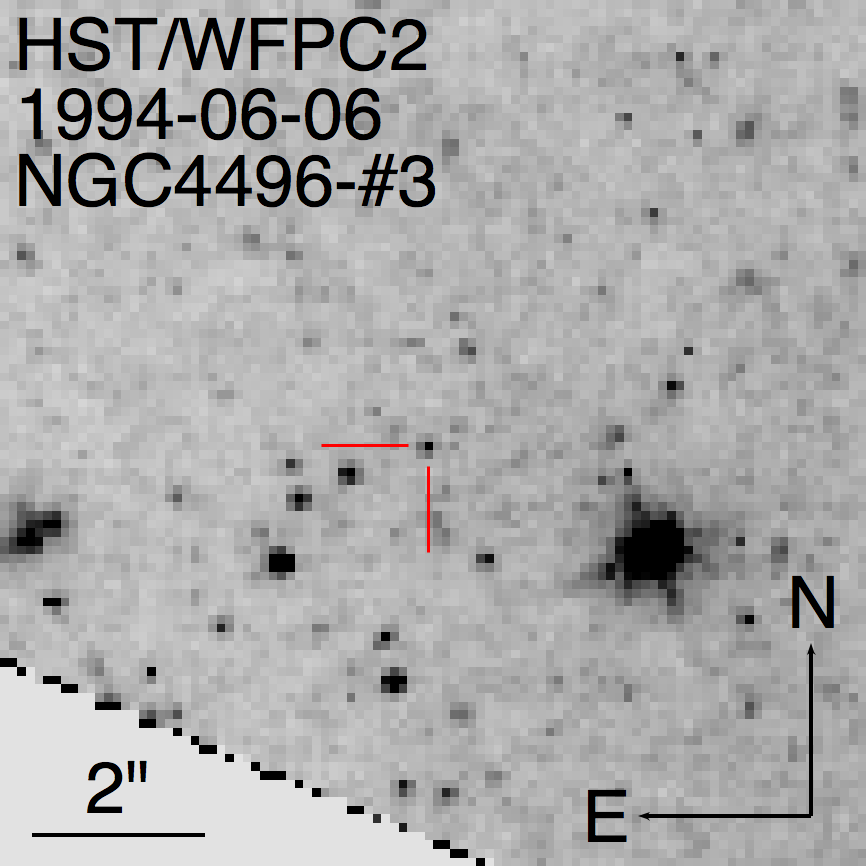}
	\includegraphics[width=0.33\linewidth,height=0.33\linewidth]{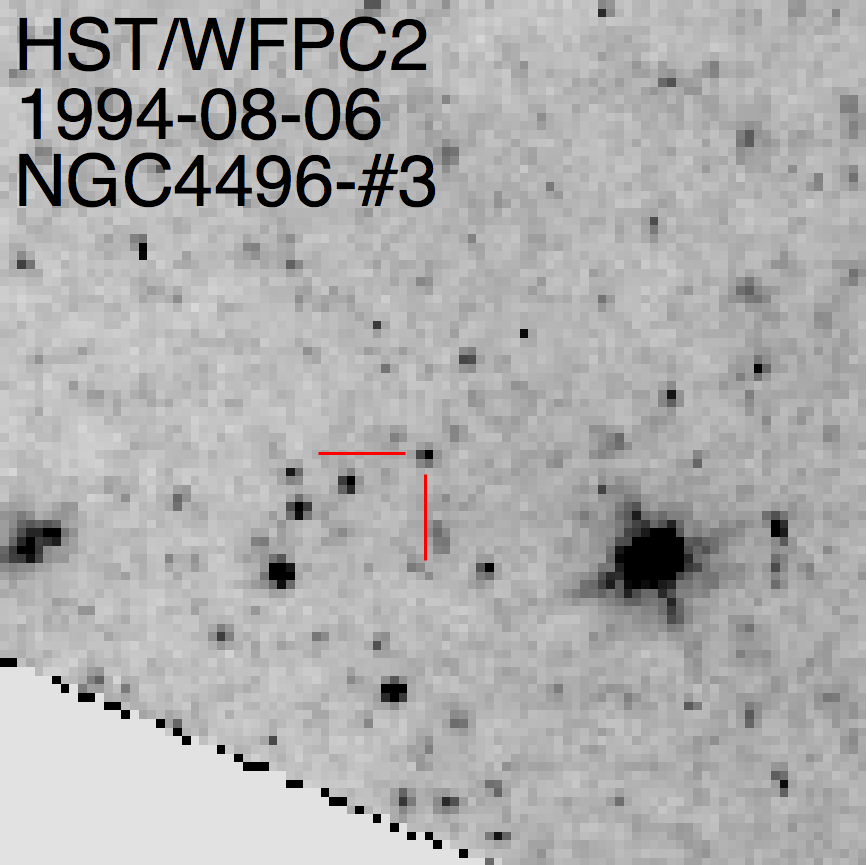}	
	\includegraphics[width=0.33\linewidth,height=0.33\linewidth]{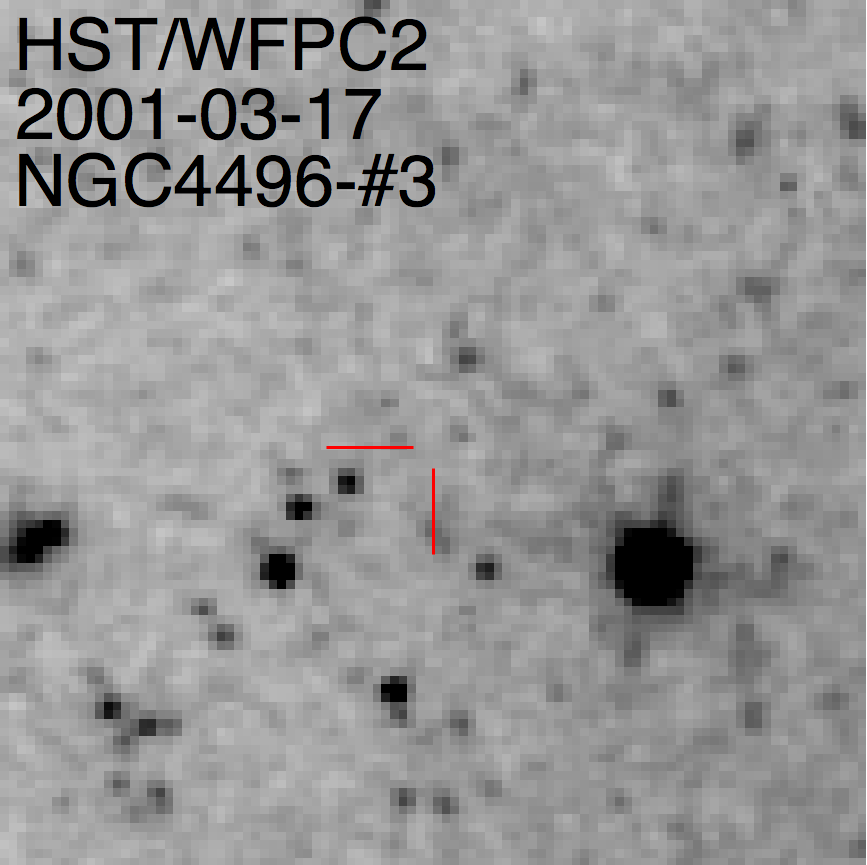}
\caption[]{{\it HST} {\it F814W} image cutouts centred on the position of NGC4496-CANDIDATE-3, similar to Fig. \ref{fig:NGC4639_1_stamps}.}
\label{fig:NGC4496_3_stamps}
\end{figure*}

The coordinates of NGC4496-CANDIDATE-3 (Fig. \ref{fig:NGC4496_3_stamps}) are 12$^{\mathrm{h}}$31$^{\mathrm{m}}$40$^{\mathrm{s}}$.72 +3\degree55\arcmin57\arcsec.2. The source lies equidistant between NGC 4496A and NGC 4496B, and is within the {\it B}=25 isophote of both galaxies, making it difficult to conclusively associate it with either. If we take the {\it B} band isophote as measuring the outer edge of the galaxy, then 65 per cent of the area of NGC 4496B is interior to the position of NGC4496-CANDIDATE-3, compared to only 16 per cent of the area of NGC 4496A.

\begin{figure}
\centering
	\includegraphics[width=0.9\linewidth]{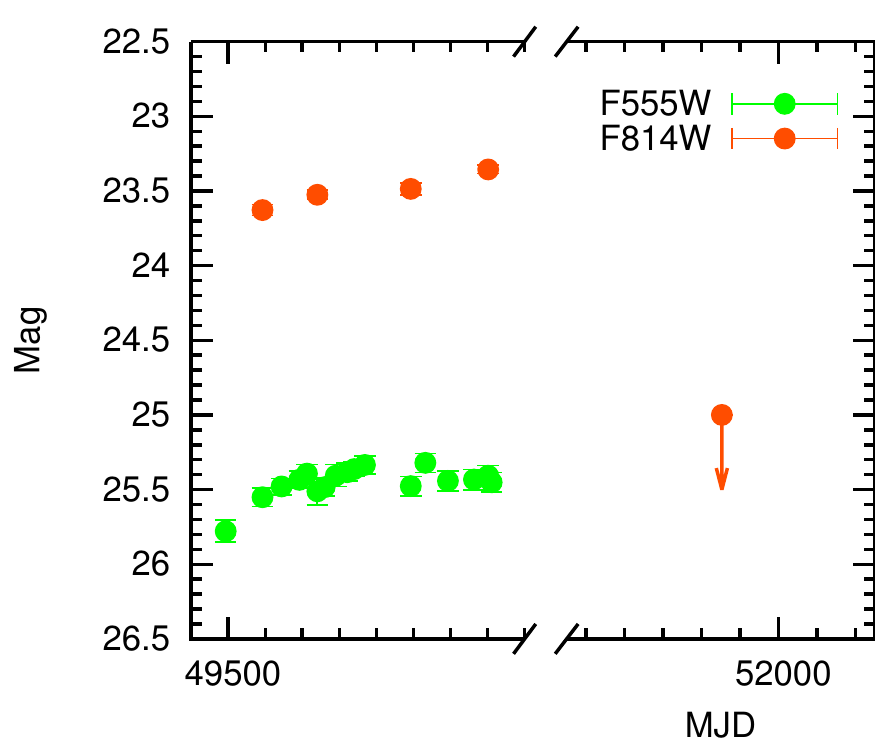}	
\caption[]{The lightcurve of NGC4496-CANDIDATE-3. Circles denote WFPC2 observations, minor tick marks are 10 d apart.}
\label{fig:NGC4496-CAND-3_LC}
\end{figure}

The photometry for NGC4496-CANDIDATE-3 is listed in Tab. \ref{tab:NGC4496-CAND-3}, while the lightcurve is shown in Fig. \ref{fig:NGC4496-CAND-3_LC}. The source appears to show variability over the $\sim$70 day period of observations in 1994, as it increases in brightness in {\it F814W} by 0.27 mag. The {\it F555W} lightcurve shows similar structure, with an increase over the period of observations in 1994, with an amplitude of $\sim$0.5 mag. The average {\it F555W} and {\it F814} mags are $25.45\pm0.10$ and $23.50\pm0.11$ mag respectively, corresponding to an absolute magnitude of {\it F555W}~$=-5.48\pm0.11$ if associated with NGC 4496A, or $-8.85$ mag if it lies in the more distant host NGC 4496B. In either case, the {\it F555W}$-${\it F814W} colour of NGC4496-CANDIDATE-3 is $1.92\pm0.15$ mag.

As for the other candidates in NGC 4496A/B, the absolute magnitude of NGC4496-CANDIDATE-3 is relatively faint for a massive RSG if it lies in NGC 4496A. However, in this case the source could be plausibly associated with NGC 4496B. If associated with NGC 4496B, then NGC4496-CANDIDATE-3 has an absolute magnitude of {\it F555W}$\sim-9$. On a Hertzsprung-Russell (HR) diagram, NGC4496-CANDIDATE-3 lies between the evolutionary tracks of a 25 and a 30 \msun\ star, as shown in Fig. \ref{fig:hr}. The colour of NGC4496-CANDIDATE-3 is also consistent with that expected for an M-type supergiant. On this basis, NGC4496-CANDIDATE-3 is consistent with a massive RSG, {\it if associated with NGC 4496B}. On the basis of the available data, it is impossible to determine whether this is indeed the case.

\begin{table}
\begin{center}
\caption{Photometry for NGC4496-CANDIDATE-3}
\begin{tabular}{lcccccccr}
\hline
Date			& Instrument	& Filter	& Mag				\\
\hline	
1994-05-27	&WFPC2/PC	& F555W	& 25.778 (0.073) \\
1994-06-06	&-			& F555W	& 25.552 (0.063) \\
			&-			& F814W	& 23.628 (0.038) \\
1994-06-11	&-			& F555W	& 25.480 (0.055) \\
1994-06-16	&-			& F555W	& 25.435 (0.059) \\
1994-06-18	&-			& F555W	& 25.393 (0.060) \\
1994-06-21	&-			& F555W	& 25.512 (0.091) \\
			&-			& F814W	& 23.525 (0.029) \\
1994-06-22	&-			& F555W	& 25.483 (0.062) \\
1994-06-26	&-			& F555W	& 25.406 (0.074) \\
1994-06-29	&-			& F555W	& 25.383 (0.060) \\
1994-07-01	&-			& F555W	& 25.362 (0.047) \\
1994-07-03	&-			& F555W	& 25.335 (0.061) \\
1994-07-16	&-			& F555W	& 25.477 (0.065) \\
			&-			& F814W	& 23.486 (0.040) \\
1994-07-20	&-			& F555W	& 25.321 (0.064) \\
1994-07-26	&-			& F555W	& 25.443 (0.068) \\
1994-08-02	&-			& F555W	& 25.433 (0.069) \\
1994-08-06	&-			& F555W	& 25.406 (0.067) \\
			&-    			& F814W	& 23.356 (0.027) \\
1994-08-07	&-			& F555W	& 25.452 (0.065) \\
2001-03-17	& WFPC2/WF4	& F814W & $<$25			\\
\hline
\end{tabular}
\label{tab:NGC4496-CAND-3}
\end{center}
\end{table}

\subsection{NGC4321-CANDIDATE-1}

\begin{figure*}
\centering
	\includegraphics[width=0.33\linewidth,height=0.33\linewidth]{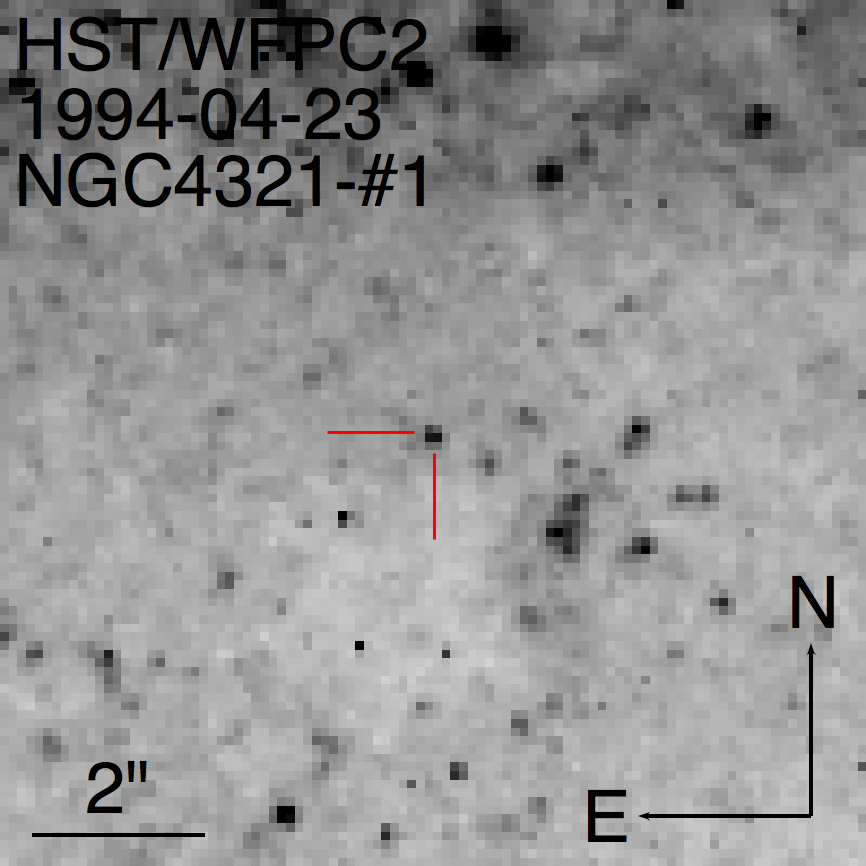}
	\includegraphics[width=0.33\linewidth,height=0.33\linewidth]{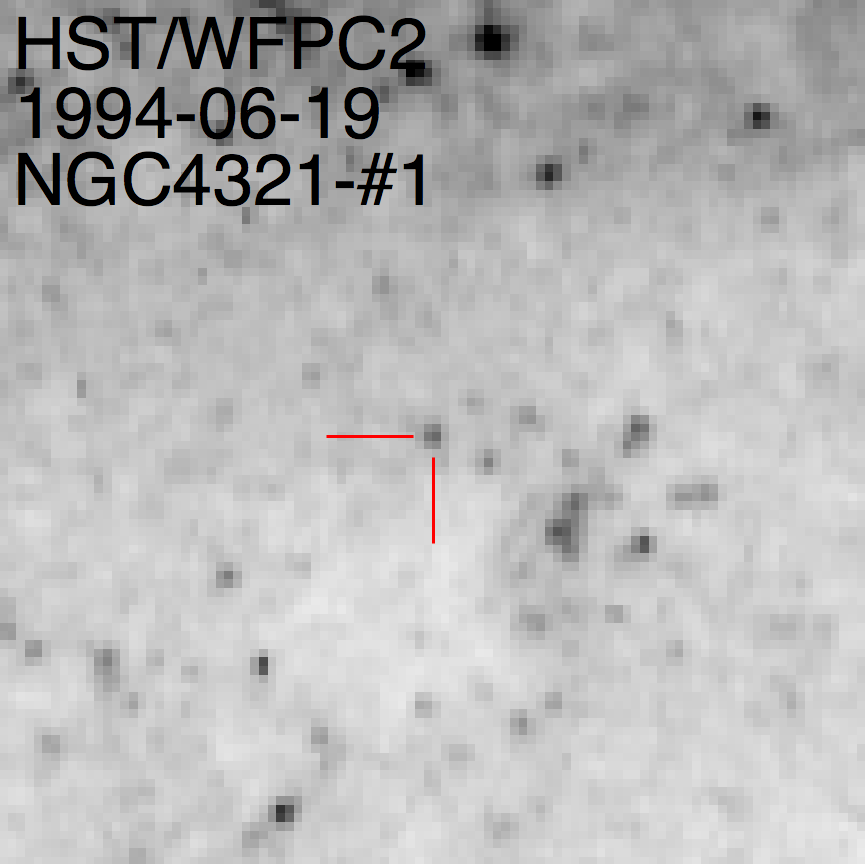}	
	\includegraphics[width=0.33\linewidth,height=0.33\linewidth]{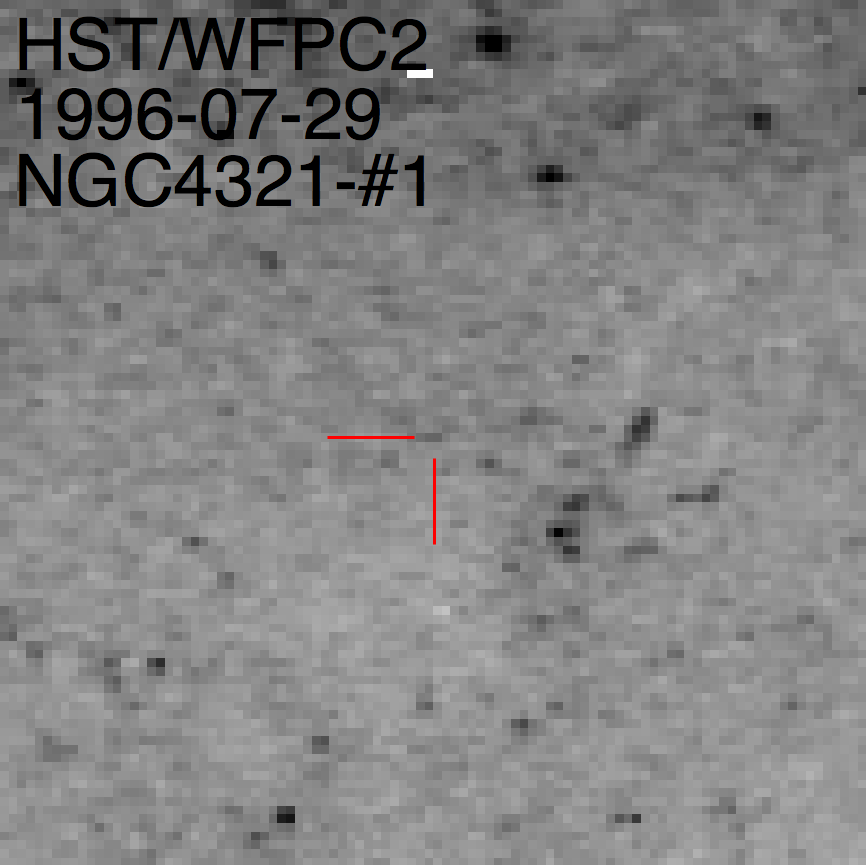}
\caption[]{{\it HST} {\it F814W} image cutouts centred on the position of NGC4321-CANDIDATE-1, similar to Fig. \ref{fig:NGC4639_1_stamps}.}
\label{fig:NGC4321_1_stamps}
\end{figure*}

\begin{figure}
\centering
	\includegraphics[width=0.9\linewidth]{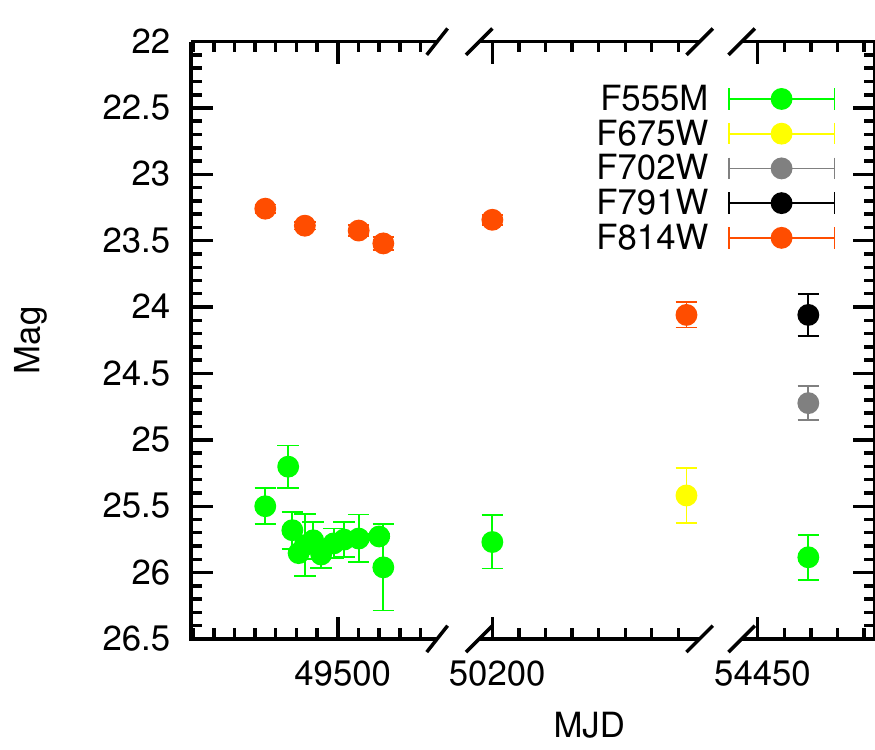}	
\caption[]{The lightcurve of NGC4321-CANDIDATE-1. Circles denote WFPC2 observations, minor tick marks are 10 d apart}
\label{fig:NGC4321-CAND-1_LC}
\end{figure}

NGC4321-CANDIDATE-1 is located at 12$^{\mathrm{h}}$22$^{\mathrm{m}}$58$^{\mathrm{s}}$.37 +15\degree47\arcmin58\arcsec.0 (Fig. \ref{fig:NGC4321_1_stamps}) in NGC 4321 (also known as M100). We adopt a distance modulus for NGC 4321 of $\upmu = 30.91 \pm 0.07$ \citep{Fre01}, and a foreground extinction of $A_\mathrm{V}=0.072$ mag. The measured magnitudes for NGC4321-CANDIDATE-1 are listed in Table \ref{tab:NGC4321-CANDIDATE-1}, while a lightcurve is plotted in Fig. \ref{fig:NGC4321-CAND-1_LC}.

The source was first observed over 57 days from April to June in 1994, when it had an average {\it F814W} magnitude of 23.40 mag. During this period the source appears to be variable, with an amplitude of 0.26 mag in {\it F814W}, which is considerably larger than the typical photometric error of $\sim$0.03 mag. The variability in {\it F555W} is even larger at 0.76 mag, although the  {\it F555W} photometry has larger uncertainties as the source is fainter in this band. Two years later in 1996 April, the source had a similar magnitude in both {\it F555W} and {\it F814W}. However, by 1994 July, the source had faded by $\sim$0.7 mag in the {\it F814W} band. A final set of observations were taken much later in 2008 January, and while the {\it F814W} filter was not used, an {\it F791W} filter image was obtained. {\it F791W} has a fairly similar bandpass to {\it F814W}, apart from having a red cut-off which is $\sim$500\AA\ shorter than that of the latter. We measured a {\it F791W} magnitude of $24.058\pm0.158$, which is identical to the {\it F814W} magnitude from 1996 July.

The absolute magnitude of NGC4321-CANDIDATE-1 is {\it F555W}~$= -5.26$, and the {\it F555W}$-${\it F814} colour is $2.29\pm0.22$ during 1994.

Based on its photometric evolution, it appears that NGC4321-CANDIDATE-1 is a variable source, rather than a star which has undergone a terminal event. From the relatively sparsely sampled lightcurve,  NGC4321-CANDIDATE-1 seems to have transitioned from a ``bright'' state in 1994, to a ``faint'' state over the following years.

\begin{table}
\begin{center}
\caption{Photometry for NGC4321-CANDIDATE-1}
\begin{tabular}{lcccccccr}	
\hline
Date			& Instrument	& Filter	& Mag		\\
\hline	
1994-04-23	& WFPC2/WF2	& F555W	& 25.500   (0.137) \\
	-		&-			& F814W	& 23.258   (0.031) \\
1994-05-04	&-			& F555W	& 25.203   (0.160) \\
1994-05-06	&-			& F555W	& 25.682   (0.140) \\
1994-05-09	&-			& F555W	& 25.854   (0.017) \\
1994-05-12	&-			& F555W	& 25.793   (0.232)\\
	-		&-			& F814W	& 23.386   (0.029) \\
1994-05-16	&-			& F555W	& 25.758   (0.138) \\
1994-05-20	&-			& F555W	& 25.863   (0.103) \\
1994-05-26	&-			& F555W	& 25.781   (0.113)  \\
1994-05-31	&-			& F555W	& 25.753   (0.133) \\
1994-06-07	&-			& F555W	& 25.743   (0.180) \\
	-		&-			& F814W  & 23.422   (0.043) \\
1994-06-17	&-			& F555W	& 25.728   (0.017) \\
	-		&-			& F555W	& 25.960   (0.325) \\
1994-06-19	&-			& F814W	& 23.521   (0.047) \\
	-		&-			& F555W	& 25.769   (0.201) \\
1996-04-27	&-			& F814W	& 23.343   (0.037) \\
1996-07-09	& WFPC2/PC	& F336W	& -				\\
	-		& -			& F439W	& -				\\
	-		& -			& F555W	& -				\\
	-		& -			& F675W	& 25.419   (0.208) 	\\
	-		& -			& F814W	& 24.058   (0.096)	\\
2008-01-04	& WFPC2/WF3	& F380W	& -				\\
	-		&-			& F439W	& -				\\
	-		&-			& F555W	& 25.885 (0.170)	\\
	-		&-			& F702W	& 24.723 (0.128)	\\
	-		&-			& F791W	& 24.058 (0.158)	\\ 
\hline
\end{tabular}
\label{tab:NGC4321-CANDIDATE-1}
\end{center}
\end{table}

\subsection{Candidates summary}

Of the six candidates for which we conducted further analysis, both NGC4639-CANDIDATE-1 and NGC4321-CANDIDATE-1 appeared to be variable. NGC4496-CANDIDATE-1 and NGC4496-CANDIDATE-2 were too faint to be consistent with massive RSGs, while NGC4496-CANDIDATE-3 is only a plausible massive star if associated with the more distant galaxy in the NGC 4496A/B pair. We hence find only a single source, NGC3021-CANDIDATE-1, which remains as a good candidate for a massive star which underwent a failed SN.

\section{Discussion}
\label{sect:discussion}

\subsection{Expected numbers}

To determine rates (or a limit to the rate) of failed SNe from the data presented in this paper is a non-trivial task. Ideally, one would quantify the detection efficiency for disappearing sources at all positions in each galaxy in our sample using artificial star tests. The results of this will vary depending on the magnitude of the source, the quality of the data, and also on the local background and degree of crowding. Once detection efficiencies have been computed, one can weight the detection probability at each position in the galaxy by the local star formation rate, and the initial mass function (and hence progenitor magnitude) given the expected mass range of failed SNe. However, such a calculation would necessitate numerous assumptions (such as the unknown magnitude range of failed SN progenitors).

As a simpler alternative, which provides a useful sanity check on our results, we simply ask the question ``Is the number of failed SN candidates seen in our sample consistent with the number of optically bright CCSNe seen in the same volume and timespan?'' This approach naturally accounts for unknowns such as the local star formation rate at any given location, and the irregular area of overlap between multiple {\it HST} fields of view. We find that of our sample, only SNe 2009hd and 2004gt lie within the area common to all our images, and exploded between the second and third epoch of imaging. If these two SNe had come from bright RSGs, which disappeared completely, then we should have detected them in our sample. However, neither SN 2004gt or SN 2009hd were detected as failed candidates in our data. SN 2004gt was a Type Ibc SN, and as deep searches for its progenitor in the same archival {\it HST} images as we have used were unsuccessful \citep{Gal05,Mau05}, it is unsurprising that we do not detect it as a failed candidate. For SN 2009hd (\cite{Eli11}), the third epoch images we used were taken of SN 2009hd itself, and as the SN is still bright in these we cannot test for the disappearance of its progenitor.

If we assume that approximately $\sim$25 per cent of massive stars may end their lives as optically dark SNe (based on the mass range for Type II SN progenitors inferred by \citealp{Sma09a}), and two bright SNe fell within our sample, then we may expect to see around $\sim$0.7 failed SNe in our sample. As discussed in Sect. \ref{sect:candidates}, only NGC3021-CANDIDATE-1 remains as a plausible failed SN candidate (along with NGC4496-CANDIDATE-3 if we adopt the larger distance for it). We hence regard the number of candidate failed SNe found as being broadly consistent with expectation, albeit with large uncertainties given the small sample size. If instead we use the results of \cite{Ger14}, who found that 30$^{+32}_{-23}$ per cent of core-collapses resulted in failed SNe, we expect to find around $\sim$0.9 failed SNe in our sample. This is again broadly consistent, albeit with uncertainties too large to say anything substantive about failed SN rates.

\begin{figure*}
\centering
	\includegraphics[width=0.9\linewidth]{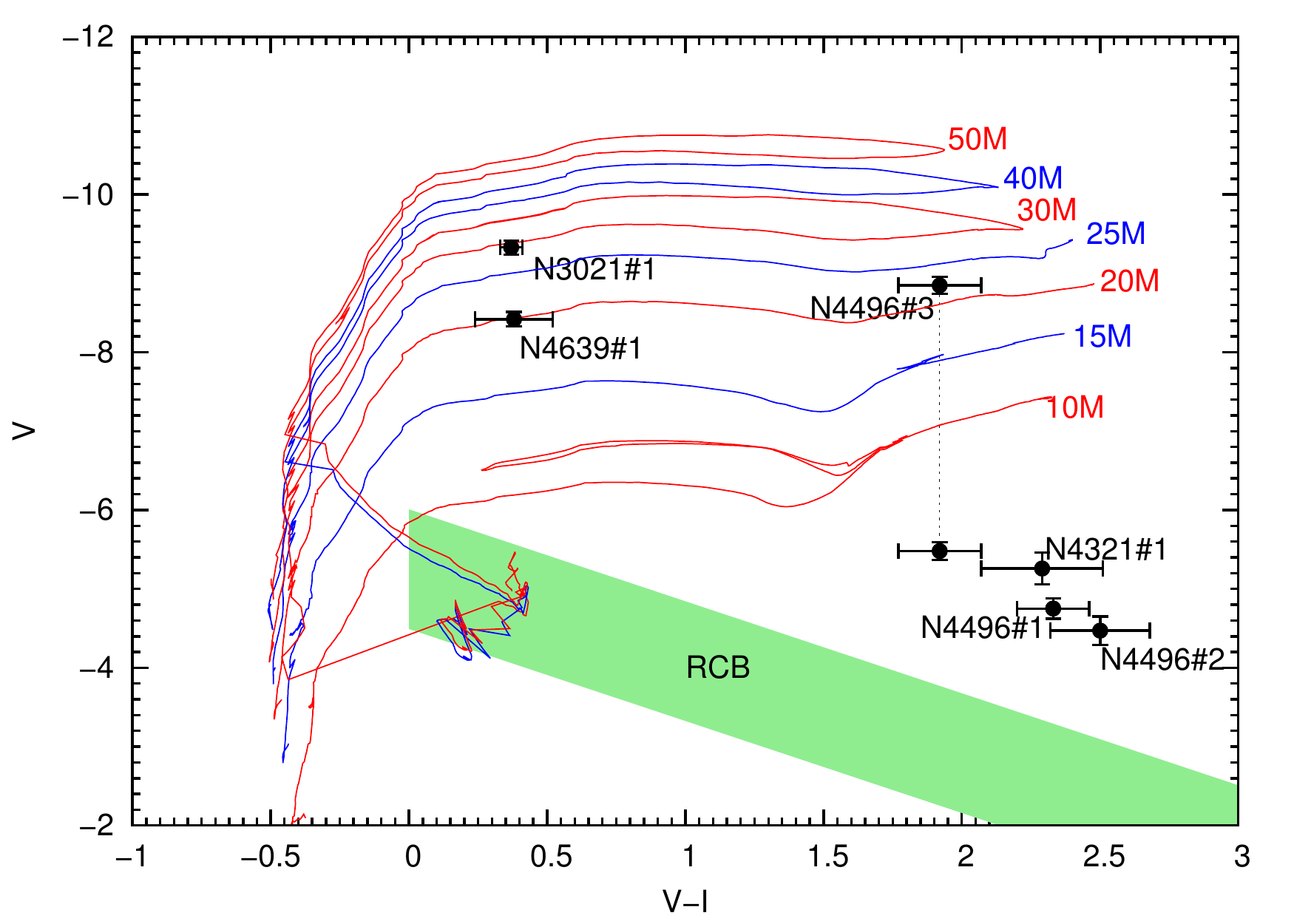}	
\caption[]{A colour magnitude diagram for the candidates presented in this paper (black points), compared to solar metallicity stellar evolutionary tracks from STARS models \protect\citep{Eld09}. The plotted magnitude for each source is the average magnitude over the first two epochs of imaging, corrected for distance and foreground reddening. The zero-age main sequence mass of each evolutionary model is indicated.  The erratic  behaviour of the tracks in the lower left is due to the more massive stars entering the Wolf-Rayet phase. The {\it F555W} and {\it F814W} magnitudes for the candidates are plotted, while the evolutionary models are in the Johnson-Cousins system, however these are sufficiently similar (with differences of a few hundredths of a magnitude) to permit a direct comparison.Two possible positions are shown for NGC4496-CANDIDATE-1, depending on whether it is associated with NGC 4496A or NGC 4496B. The green shaded region shows the range of RCB star magnitudes from \protect\cite{Tis09}.}
\label{fig:hr}
\end{figure*}

\subsection{What do optically dark SNe really look like?}

While the premise of this paper was to search for the disappearance of a massive and luminous RSG without an accompanying optical display, it is worth considering in more detail whether {\it any} transient would be associated with such an event. There have been some attempts to answer this question from a theoretical basis \citep[e.g.][]{Nad80,Lov13,Pir13}, which have focused on the effect of removing around 0.1\msun\ of mass from the core of a massive star via neutrino emission. The outer layers of the star will adjust to this rapid loss of mass, and in the models of \cite{Lov13} a fraction of the envelope is ejected with a kinetic energy of about $10^{47}$~ergs. This would be accompanied by a faint red transient which lasts some hundreds of days. \cite{Pir13} suggested that there would be shock breakout emission from such an event, peaking in the ultraviolet (UV) and lasting only a few days.

A transient with brief, UV bright emission would likely not be seen in our data, as our typical cadence is extremely low. However, a slow red transient would most likely be seen in our data. \cite{Lov13} calculated that such a transient would have a luminosity of $10^{39}$ -- $10^{40}$~ergs~s$^{-1}$. At a typical distance of $\sim$20 Mpc, this would have an apparent magnitude of {\it R}$\sim$20. While such a transient may well be missed in ground based surveys with 1-m or 2-m class telescopes, which typically have a limiting magnitude of $\gtrsim 21$ (or even shallower in the case of the LOSS survey, which had a limiting magnitude of 18--19.5 mag; \citealp{Lea11}), it would be clearly visible in our {\it HST} data.

\subsection{Contaminants}

A major concern for any attempt to find optically dark SNe is that variable stars or other transients could mimic the disappearance of a massive star, either through appearing brighter in the first two epochs of imaging, or fainter in the third epoch.

R Coronae Borealis (RCB) stars are one such possible contaminant. These are evolved, hydrogen poor F or G supergiants, which experience repeating, irregular episodes where they decline by up to 8 magnitudes for hundreds of days \citep{Cla96}. This is thought to be caused by the formation of clouds of amorphous carbon dust which can temporarily obscure the star. Related to RCBs are DY Per type stars, which are cooler than typical RCBs with effective temperatures of $\sim$3500 K. RCBs are quite rare, with only $\sim$50 known in the Milky Way, while DY Per type stars are even less common. The absolute magnitude of RCBs are, however, fainter than the candidates found in our sample. The region occupied by known RCB stars from \cite{Tis09} is shown in Fig. \ref{fig:hr}, and it is clear that all of our candidates lie above this region. We hence discount RCBs as a possible explanation for our candidates.

\begin{figure}
\centering
	\includegraphics[width=0.7\columnwidth,angle=270]{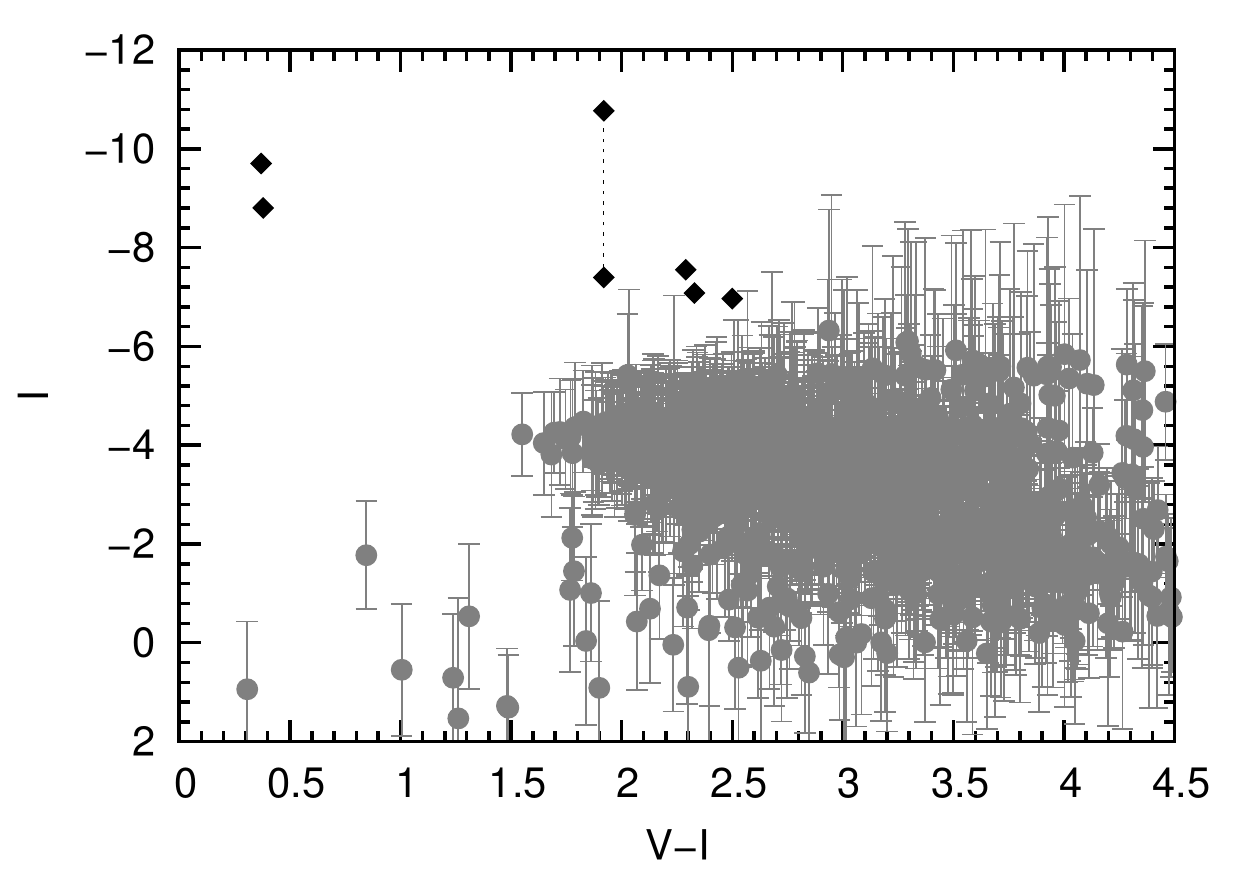}	
\caption[]{Colour magnitude diagram for Mira variables discovered in the LMC by the OGLE survey \protect\citep{Sos09, Uda08}. The mean {\it I}-band magnitudes of the OGLE Miras are plotted as grey points, with error bars corresponding to their amplitude. The failed SN candidates discussed in this paper are shown as black diamonds.}
\label{fig:mira}
\end{figure}

Pulsating stars, which lie in the instability strip of the HR diagram are another source of concern. Some, such as the RV Tauri variables, can be dismissed as being too faint to appear in our data; even the brightest R Scuti has a luminosity log~L/\lsun $<$4 dex. However, some long period variables, and in particular Mira variables, are more of a concern. Miras  are pulsating, late type stars which can have amplitudes of several magnitudes, and periods of hundreds of days. To assess whether any of our candidates could be Miras, in Fig. \ref{fig:mira} we compare them to a sample of Miras found in the Large Magellenic Cloud by the OGLE survey \citep{Sos09, Uda08}. The population of Mira type variables in the LMC appears to be fainter, and for the most part significantly redder than our candidates.
 
Finally, we consider the possibility that known classes of transients could mimic a disappearing RSG. This would be more of a concern if we had only two detections before a source is seen to disappear. In such a scenario, the first and second detections of a source could be two successive outbursts, perhaps of a recurrent nova, or of an S-Doradus type luminous blue variable, which happened to be observed at a similar magnitude. In the third image, the hypothetical source could appear to have disappeared as it is not in outburst. However, as each of the candidates in our sample are detected on more than six separate occasions before disappearing, we regard this as unlikely.

\subsection{Was an optically dark SN seen in our sample?}

We cannot conclusively say whether an optically dark SN was seen in our sample. Our best candidate, NGC3021-CANDIDATE-1 appears to lie in a complex of bright sources in a star forming arm of a massive spiral galaxy. On the HR diagram shown in Fig. \ref{fig:hr}, NGC3021-CANDIDATE-1 lies close to the track for a 30\msun\ star. The candidate lies above the 25\msun\ track, and so we adopt a mass of 25--30 \msun. In the STARS models plotted, if the star lies at the upper end of this mass range, then it may not end its life as an RSG, but rather cross the HR diagram to the blue to become a Wolf Rayet star.

The colour of NGC3021-CANDIDATE-1 is bluer than that of an RSG, and is in fact consistent with an F8 supergiant \citep{Pic98}. It is important to note however, that the effective temperature of a star at the point it explodes is dependent on its radius and the extent of any mass loss. In the case of SN 2011dh, a yellow supergiant (YSG) was observed to explode as a core-collapse SN \citep{Mau11,Van11}, and while the progenitor of SN 2011dh likely had its envelope stripped by a binary companion \citep{Fol14}, it at least establishes a precedent for massive stars exploding as YSGs. Regardless of the progenitor temperature, the progenitor luminosity should scale monotonically with core (and hence progenitor) mass.

We checked all available online databases for any reported SN or transient at the position of NGC3021-CANDIDATE-1 between 2006 and 2013. NGC 3021 was targeted by the LOSS survey \citep{Lea11}, and was also within the footprint of PanSTARRS-1, PTF and CRTS. To the best of our knowledge, none of these surveys reported a SN or a candidate transient at the location of NGC3021-CANDIDATE-1.

While NGC4496-CANDIDATE-3 is consistent with a massive RSG if associated with the more distant galaxy NGC 4496B, it is impossible to determine this on the basis of the available data. Furthermore, if associated with NGC 4496B, then the greater distance of the host would formally exclude this from our volume limited sample.

\section{Conclusions}
\label{sect:conclusions}

{\it HST} imaging has been shown to be a powerful tool in the search for disappearing massive stars. In our preliminary study we have found one candidate failed SNe which merits further investigation. {\it HST} delivers an order of magnitude improvement in spatial resolution over ground based data. We note that many of our candidates would have been blended with nearby sources in ground based images, rendering their identification more difficult.

With the successful demonstration of image subtraction using archival {\it HST} images, one can begin to consider designing an experiment to increase the number of candidates seen. We note that while we required galaxies which had been observed three times with {\it HST}, there are over one hundred more galaxies which had two available epochs of imaging. Approximately 30 of these have some overlap between the first and second epoch, hence if one were to obtain a third epoch of tailored observations for each of these, our sample size would be increased by a factor of $\gtrsim$2.  

We estimate that to monitor a galaxy would require an image every 4 months, ideally in two filters ({\it F555W} and {\it F814W}). A 4 month cadence would ensure that faint, but relatively long-lived transients such as those modelled by \cite{Lov13} would not be missed, and would also permit long period variable stars such as Miras to be identified. In a single orbit with {\it HST}+ACS one can reach limiting magnitudes of $V=27.9$ and $I=27.4$; at a distance of 20 Mpc this corresponds to a limiting absolute magnitude of $V=-3.6$ and $I=-4.1$. So, to monitor a sample of 20 galaxies to this depth one would require 120 orbits of {\it HST} time per year, a significant investment of time. The probability of success could be increased by selecting galaxies which have a relatively prolific SN rate, such as NGC 4303 which has hosted four CCSNe in the last 16 years. Such a survey would also facilitate a wide range of science, including monitoring the variability of massive stars, and searching for a range of other faint transients such as stellar mergers, common-envelope ejections and luminous red novae (i.e. the ``gap transients'' which lie between novae and supernovae; \citealp{Kul07}).

Our original survey was designed to find failed SNe with only three epochs of data -- two early epochs in which a star was present, and a late epoch in which the same star had vanished. However, the candidates presented in this work appear to have been preferentially identified in galaxies for which significantly more data were available. This suggests that the original expectation that three epochs would be sufficient to find failed SNe may be overly optimistic, although with deep followup observations we could still verify many candidates. Running an ongoing {\it HST} survey with higher cadence monitoring would help to separate out failed SNe from contaminants.

Finally, we note that alongside the ongoing observational project to find failed SNe and their progenitors, there is a pressing need for theoretical work on the mass ranges of stars which are expected to explode, and to understand where the islands of ``explodability'' (e.g. \cite{Ugl12};  \cite{Pej14}; \cite{Ert15}) may lie.

\section*{acknowledgments}

We thank the anonymous referee for a careful reading of the paper, and for constructive and helpful comments. We thank Seppo Mattila, Stephen Smartt and John Eldridge for useful discussions and suggestions on this work.

This work was supported by the European Union FP7 programme through ERC grant number 320360.

Based on observations made with the NASA/ESA Hubble Space Telescope, and obtained from the Hubble Legacy Archive, which is a collaboration between the Space Telescope Science Institute (STScI/NASA), the Space Telescope European Coordinating Facility (ST-ECF/ESA) and the Canadian Astronomy Data Centre (CADC/NRC/CSA). Some of the data presented in this paper were obtained from the Mikulski Archive for Space Telescopes (MAST). STScI is operated by the Association of Universities for Research in Astronomy, Inc., under NASA contract NAS5-26555. This research has made use of the HyperLeda database (http://leda.univ-lyon1.fr) and of the NASA/IPAC Extragalactic Database (NED) which is operated by the Jet Propulsion Laboratory, California Institute of Technology, under contract with the National Aeronautics and Space Administration.

\appendix
\section{Archival data}

\begin{table*}
	\begin{minipage}{170mm}
		\caption{Archival images used for each galaxy.  $\Delta t$ is  the period (in years) between a given epoch and the earliest observations of that galaxy.}
		\begin{tabular}{lllllr}
			\hline		
			
			~ & Date & Instrument & Exposure (s)  & $\Delta t$ (yr) & HLAfilename or Dataset		\\
			
			\hline
			
			NGC 1058 & & & &  & \\
			Epoch 1:  & 2001-07-03 		& WFPC2   	& 460   	   	& -  		& hst\_09042\_n2\_wfpc2\_f814w\_wf\_drz.fits  	 	\\ 
			Epoch 2:  	& 2007-10-29 		& WFPC2  	& 320 	   	& 6.33 	& hst\_11119\_01\_wfpc2\_f814w\_wf\_drz.fits 		\\ 
			Epoch 3:  	& 2008-11-20 		& WFPC2  	& 700   	  	& 7.39 	& hst\_10877\_32\_wfpc2\_f814w\_wf\_drz.fits 		\\
			&				&			&		&				&										\\
			
			NGC 1566 & & & &  & \\
			Epoch 1:  & 1995-08-30 		& WFPC2   	& 300   	   	& -  		& 1995-08-30\_mosaic\_NGC1566.fits\footnotemark[1]   			\\ 
			Epoch 2:  & 2009-03-11 		& WFPC2  	& 600   	   	& 13.54 	& hst\_11987\_15\_wfpc2\_f814w\_wf\_drz.fits 		\\ 
			Epoch 3:  & 2013-09-03 		& WFC3   		& 989     	& 18.02 	& hst\_13364\_17\_wfc3\_uvis\_f814w\_drz.fits 						\\
			&				&			&		&				&										\\
			
			NGC 2841 & & & &  & \\
			Epoch 1:  & 2000-02-29  					&		&		&		&										\\
			& --- 2000-04-09 	& WFPC2   	& 44000       	& -  		& hlsp\_sgal\_hst\_wfpc2\_n2841\_f814w\_v1\_mosaic-sci\_drz.fits   \\
			Epoch 2:  & 2005-04-05 		& WFPC2 	 & 500   	    	& 4.6 	& hst\_10402\_12\_wfpc2\_f814w\_wf\_drz.fits 		\\ 
			Epoch 3:  & 2010-01-05 		& WFC3   		& 1412   	    	& 9.36 	& hst\_11360\_r2\_wfc3\_uvis\_f814w\_drz.fits 		\\ 
			&				&			&		&				&										\\
			
			NGC 3021 & & & &  & \\
			Epoch 1:  	& 2005-11-12 		& ACS/WFC  	& 4800   	  		& -  		& HST\_10497\_13\_ACS\_WFC\_F814W\_drz.fits   	\\ 
			Epoch 2:  	& 2006-01-02 		& ACS/WFC  	& 4800   	  		& 0.14 	& HST\_10497\_24\_ACS\_WFC\_F814W\_drz.fits 	\\ 
			Epoch 3:  	& 2013-02-07 		& WFC3   		& 1126   	 		& 7.64 	& hst\_12880\_76\_wfc3\_uvis\_f814w\_drz.fits 					\\ 
			&				&			&		&				&										\\
			NGC 3344 & &  & & & \\
			Epoch 1:  & 2001-11-13 		& WFPC2   	& 460   	   	& -  		& hst\_09042\_40\_wfpc2\_f814w\_wf\_drz.fits   	\\ 
			Epoch 2:  & 2012-02-24 		& ACS/WFC  	& 900      	& 10.29 	& jbt10k020\_drz.fits 							\\ 
			Epoch 3:  & 2014-03-02 		& WFC3   		& 980  	    	& 12.31 	& hst\_13364\_22\_wfc3\_uvis\_f814w\_drz.fits  						\\ 
			&			&	&		&			&							\\
			\multicolumn{6}{l}{NGC 3368 - Position } \\
			Epoch 1:	& 1994-04-06 		& WFPC2		& 3000			& -		& hst\_05415\_01\_wfpc2\_f814w\_wf.fits			\\
			Epoch 2: 	& 1995-04-06 		& WFPC2		& 3000			& 1.00	& hst\_05415\_0p\_wfpc2\_f814w\_wf.fits			\\
			Epoch 4:  & 2009-11-12 		& WFC3   		& 350 	   	& 15.60  	& hst\_11646\_02\_wfc3\_uvis\_f814w\_drz.fits  	\\
			&			&	&		&			&							\\
			\multicolumn{6}{l}{NGC 3368 - Position 2} \\
			Epoch 1:  & 2000-12-12 		& WFPC2 	& 700		& - 		& hst\_08602\_56\_wfpc2\_f814w\_wf\_drz.fits		\\  
			Epoch 2:  & 2001-11-06  		& WFPC2 	& 320 		& 0.90 	& hst\_09042\_n5\_wfpc2\_f814w\_wf\_drz.fits  		\\  
			Epoch 4:  & 2009-11-12 		& WFC3   		& 350 	   	& 8.93  	& hst\_11646\_02\_wfc3\_uvis\_f814w\_drz.fits  	\\
			&				&	&		&		&																\\		
			NGC 3370 & & & &  & \\
			Epoch 1: 	& 2001-07-06 		& WFPC2 	& 460		& - 		& hst\_09042\_41\_wfpc2\_f814w\_wf\_drz.fits	 	\\  
			Epoch 2: 	& 2003-03-31  		&			&				&		&										\\
			& -- 2003-05-22 	& ACS/WFC 	& 24000 	 		& 1.73 	& HST\_MOS\_1007\_ACS\_WFC\_F814W\_drz.fits  \\ 
			Epoch 3:  & 2013-01-06  		& WFC3  		& 1136 	  	& 11.50	& hst\_12880\_77\_wfc3\_uvis\_f814w\_drz.fits							 \\
			&			&	&		&			&							\\
			
			NGC 3627 & &   & & \\
			Epoch 1: 	& 2001-02-24 		& WFPC2 	& 700		& - 		& hst\_08602\_29\_wfpc2\_f814w\_wf\_drz.fits	 \\  
			Epoch 2: 	& 2001-05-28  		& WFPC2 	& 700 		& 0.25 	& hst\_08602\_30\_wfpc2\_f814w\_wf\_drz.fits  \\  
			Epoch 3:  	& 2009-12-14  		& ACS  		& 520 	  	& 8.81  	& HST\_11575\_01\_ACS\_WFC\_F814W\_drz.fits  \\
			&				&			&		&		&									\\
			NGC 3982 & & & &  & \\
			Epoch 1: 	& 2000-04-02 		& WFPC2 	& 5000 			& - 		& hst\_08100\_02\_wfpc2\_f814w\_wf\_drz.fits	 \\  
			Epoch 2: 	& 2001-10-06  		& WFPC2 	& 460 	 		& 1.51 	& hst\_09042\_o5\_wfpc2\_f814w\_wf\_drz.fits  \\  
			Epoch 3:  	& 2009-12-14  		& WFC3  		& 700	   	& 9.35  	& hst\_11570\_22\_wfc3\_uvis\_f814w\_drz.fits  \\
			&				&			&		&		&								\\
			
			\hline
			\multicolumn{6}{l}{\footnotemark[1]{Mosaic created manually as discussed in the text.} }
			
		\end{tabular}
		\label{tab:hla_data_1}	
	\end{minipage}
\end{table*}

\begin{table*}
	\begin{minipage}{170mm}
		\caption{Continuation of archival images used}
		\begin{tabular}{lllllr}
			
			\hline
			
			~ & Date & Instrument & Exposure (s)  & $\Delta t$ (yr) & HLAfilename or Dataset		\\
			
			\hline
			
			NGC 4038 & & & & &  \\
			Epoch 1: 	& 1996-01-20 		& WFPC2 	& 2120 			& - & hst\_05962\_01\_wfpc2\_f814w\_wf\_drz.fits	 \\  
			Epoch 2: 	& 2004-07-21  		& ACS/WFC 	& 1680 			& 8.51 & HST\_10188\_10\_ACS\_WFC\_F814W\_drz.fits  \\  
			Epoch 3:  	& 2010-01-22  		& WFC3  		& 1032 	  	& 14.02  & hst\_11577\_20\_wfc3\_uvis\_f814w\_drz.fits  \\
			&				&			&		&			&							\\
			
			NGC 4321 & & & & &  \\
			Epoch 1: 	& 1994-04-23 		& WFPC2 	& 3600 		& - & hst\_05397\_13\_wfpc2\_f814w\_wf\_drz.fits	 \\  
			Epoch 2: 	& 1994-06-19  		& WFPC2 	& 3600 	 	& 0.16 & hst\_05397\_12\_wfpc2\_f814w\_wf\_drz.fits  \\  
			Epoch 3:  	& 1996-07-29  		& WFPC2  	& 1210 	   & 2.27  & hst\_06584\_02\_wfpc2\_f814w\_wf\_drz.fits  \\
			&				&	&		&			&							\\
			
			NGC 4414 & & & &  & \\
			Epoch 1: 	& 1995-04-06 		& WFPC2 	& 2500 	 & - & hst\_05397\_27\_wfpc2\_f814w\_wf\_drz.fits	 \\  
			Epoch 2: 	& 1995-06-10  		& WFPC2 	& 2500 	 & 0.18 & hst\_05397\_2i\_wfpc2\_f814w\_wf\_drz.fits  \\  
			Epoch 3:  	& 1999-04-29  		& WFPC2  	& 6400 	  & 4.07  & hlsp\_sgal\_hst\_wfpc2\_n4414-1\_f814w\_v1\_mosaic-sci\_drz.fits  \\
			&				&	&		&		&								\\ 
			
			NGC 4496 & & & & &  \\
			Epoch 1: 	& 1994-06-06 		& WFPC2 	& 4000  & - & hst\_05427\_02\_wfpc2\_f814w\_wf\_drz.fits	 \\  
			Epoch 2: 	& 1994-08-06  		& WFPC2 	& 4000  & 0.21 & hst\_05427\_5c\_wfpc2\_f814w\_wf\_drz.fits  \\  
			Epoch 3:  	& 2001-03-17  		& WFPC2  	& 640  & 6.78  & hst\_08599\_64\_wfpc2\_f814w\_wf\_drz.fits  \\
			&				&	&		&			&							\\
			NGC 4639 & & & &  & \\
			Epoch 1: 	& 1995-05-17 		& WFPC2 	& 7800 	 & - & hst\_05981\_04\_wfpc2\_f814w\_wf\_drz.fits	 \\  
			Epoch 2: 	& 1995-06-03  		& WFPC2 	& 5200 	 & 0.047 & hst\_05981\_08\_wfpc2\_f814w\_wf\_drz.fits  \\  
			Epoch 3:  	& 2009-08-07  		& WFC3  		& 800 	 & 14.24  & hst\_11570\_23\_wfc3\_uvis\_f814w\_drz.fits  \\
			&				&			&		&			&							\\
			NGC 4651 & & & &  & \\
			Epoch 1: 	& 1994-05-20 		& WFPC2 	& 660 	 		& - 		& hst\_05375\_09\_wfpc2\_f814w\_wf\_drz.fits	 \\  
			Epoch 2: 	& 2007-04-27 		& WFPC2 	& 1200 	 		& 12.9 	& hst\_10803\_03\_wfpc2\_f814w\_wf\_drz.fits  \\  
			Epoch 3:  	& 2010-11-21  		& ACS   		& 1090 	  	& 16.5  	& jbks01020\_drz.fits 			 \\ 
			&				&			&		&				&							\\

			\hline
			
		\end{tabular}
		\label{tab:hla_data_2}	
	\end{minipage}
\end{table*}

\clearpage

\end{document}